\documentclass[sigconf]{acmart}

\usepackage{siunitx}

\AtBeginDocument{%
  \providecommand\BibTeX{{%
    \normalfont B\kern-0.5em{\scshape i\kern-0.25em b}\kern-0.8em\TeX}}}

\copyrightyear{2024}
\acmYear{2024}
\setcopyright{rightsretained}
\acmConference[UIST '24]{The 37th Annual ACM Symposium on User Interface Software and Technology}{October 13--16, 2024}{Pittsburgh, PA, USA}
\acmBooktitle{The 37th Annual ACM Symposium on User Interface Software and Technology (UIST '24), October 13--16, 2024, Pittsburgh, PA, USA}
\acmDOI{10.1145/3654777.3676440}
\acmISBN{979-8-4007-0628-8/24/10}

\begin{document}

\title{JetUnit: Rendering Diverse Force Feedback in Virtual Reality Using Water Jets}

\author{Zining Zhang}
\affiliation{%
  \institution{University of Maryland}
  \city{College Park}
  \state{MD}
  \country{USA}
}
\email{znzhang@umd.edu}

\author{Jiasheng Li}
\affiliation{%
  \institution{University of Maryland}
  \city{College Park}
  \state{MD}
  \country{USA}
}
\email{jsli@umd.edu}

\author{Zeyu Yan}
\affiliation{%
  \institution{University of Maryland}
  \city{College Park}
  \state{MD}
  \country{USA}
}
\email{zeyuy@umd.edu}

\author{Jun Nishida}
\affiliation{%
  \institution{University of Maryland}
  \city{College Park}
  \state{MD}
  \country{USA}
}
\email{jun@umd.edu}

\author{Huaishu Peng}
\affiliation{%
  \institution{University of Maryland}
  \city{College Park}
  \state{MD}
  \country{USA}
}
\email{huaishu@umd.edu}

\renewcommand{\shortauthors}{Zhang et al.}

\begin{abstract}
We propose JetUnit, a water-based VR haptic system designed to produce force feedback with a wide spectrum of intensities and frequencies through water jets. 
The key challenge in designing this system lies in optimizing parameters to enable the haptic device to generate force feedback that closely replicates the most intense force produced by direct water jets while ensuring the user remains dry. 
In this paper, we present the key design parameters of the JetUnit wearable device determined through a set of quantitative experiments and a perception study. 
We further conducted a user study to assess the impact of integrating our haptic solutions into virtual reality experiences. The results revealed that, by adhering to the design principles of JetUnit, the water-based haptic system is capable of delivering diverse force feedback sensations, significantly enhancing the immersive experience in virtual reality.
\end{abstract}

\ccsdesc[500]{Human-centered computing~Haptic devices}
\ccsdesc[500]{Human-centered computing~Human computer interaction (HCI)}
\ccsdesc[500]{Human-centered computing~Interaction devices}

\keywords{haptics, water jets, force feedback, VR}

\maketitle

\begin{figure}[h]
  \centering
  \includegraphics[width=\linewidth]{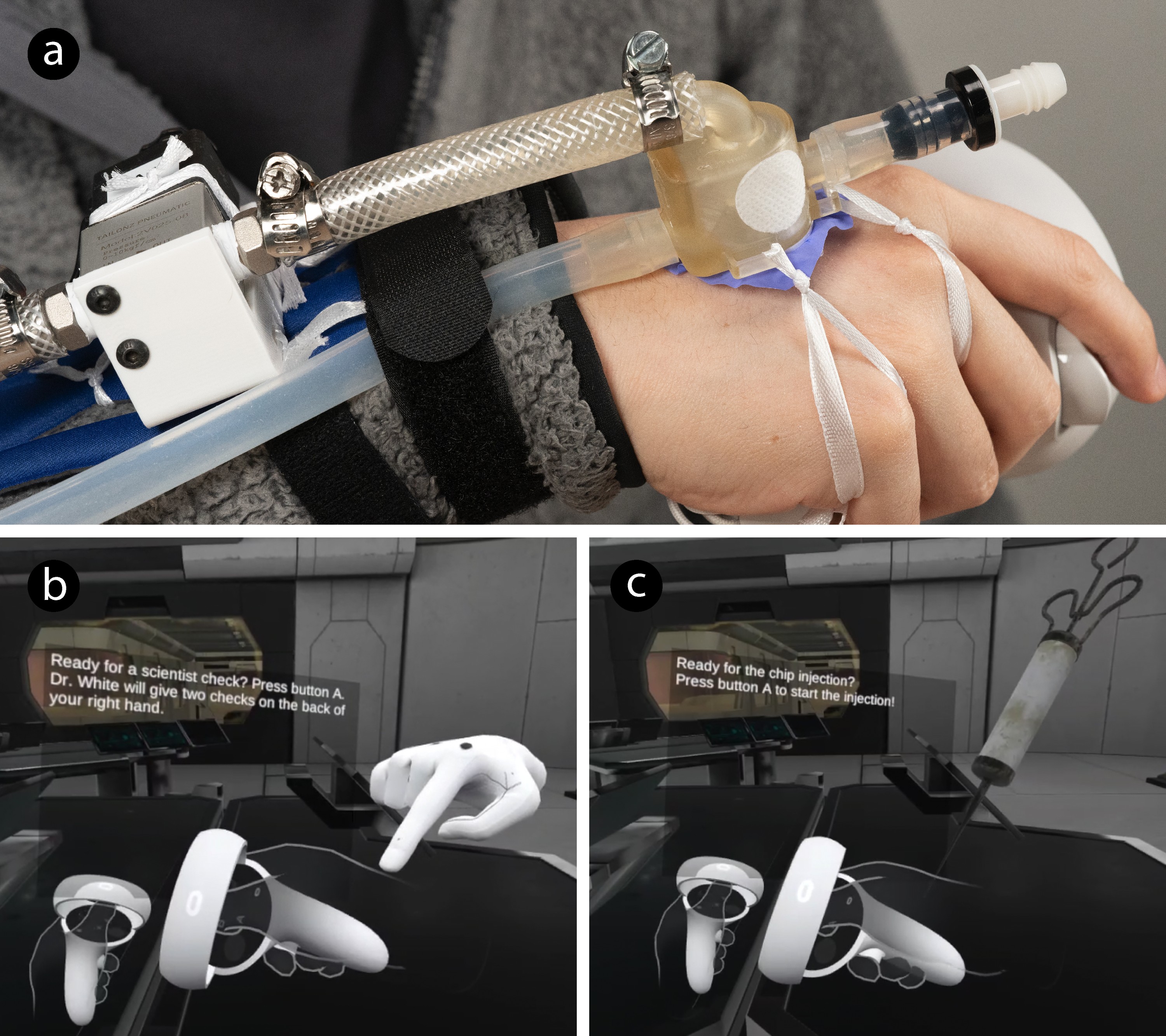}
  \caption{(a) The JetUnit offers force feedback over a wide range of perceived force intensities and pulsing frequencies. A single wearable JetUnit can render interactions ranging from (b) a gentle touch to (c) a progressively accelerating injection.}
  \Description{Figure 1 (a) shows the JetUnit wearable device, which is able to render both gentle touch as shown in (b), and (c) progressively accelerating injection.}
  \label{fig:teaser}
\end{figure}

\section{Introduction}
Studies have shown that realistic force feedback can significantly enhance VR immersiveness \cite{richard1996effect}.
Recent research has explored various mechanisms for delivering force feedback in VR, including pneumatic systems \cite{kanjanapas2019design, gunther2020pneumovolley, delazio2018force, choi2022design, he2015pneuhaptic, talhan2023soft}, exoskeletons \cite{schorr2017fingertip, tsai2019elasticvr, tsai2022impactvest, teng2021touch, tsai2019elastimpact, pezent2022design, jouybari2021cogno}, electric muscle stimulation (EMS) \cite{farbiz2007electrical, hosono2022feedback, kurita2016stiffness, lopes2017providing}, and combinations thereof \cite{nith2021dextrems, 10.1145/2807442.2807443}.
While each of these approaches can effectively deliver certain types of force feedback, they are often limited by the strength, frequency, or pattern of the force that can be produced, thus constraining the range of VR application scenarios. 
For example, a pneumatic system may excel at simulating gentle touches \cite{talhan2023soft}, but may fall short in replicating the sensation of a sudden and intense impact. 
On the other hand, a rubber-based haptic device \cite{tsai2022impactvest} may be adept at conveying instant impacts, but struggle with soft, gentle touches or rapid, repeated sensations, such as the feeling of continuous raindrops. 
Therefore, achieving diverse force feedback that matches various user scenarios remains challenging.

In this paper, we propose \textit{JetUnit}, a working wearable prototype engineered to provide force feedback across a broad spectrum of perceived force intensities and frequencies for various VR scenarios. 
Central to our system is the use of water jets to render force feedback. 
Compared with other mediums such as air, water was chosen due to its incompressible nature \cite{10.1145/3472749.3474751, han2018hydroring}. According to the Navier-Stokes equation, a pressure change in incompressible fluids can directly result in a change in velocity and thus impact force, allowing for more efficient momentum transfer. %
This property allows for the delivery of both strong and gentle force feedback in a variety of patterns on the user's body, akin to the experience of with water-based massage systems like Jacuzzis. 
In the meantime, the JetUnit distinguishes itself as a self-contained wearable system---while offering force feedback, the JetUnit ensures users remain dry, making it suitable for VR applications. 

The key to JetUnit implementation is its custom-designed chamber unit. 
The chamber unit propels water directly onto a thin membrane, which transmits haptic sensations to the users' skin.
The membrane, securely sealed at the chamber's opening, ensures the water remains contained. 
However, this design risks reducing the intensity of the water streams due to the accumulation of water inside the chamber and the turbulence introduced thereafter. 
To address this, we have implemented four measures in the chamber design. 
The first measure involves connecting the outlet of the chamber to a recycling pump to facilitate efficient water drainage.
The second measure is adding a ring channel with side openings adjacent to the membrane sealing area. This design enables rapid evacuation of water from the chamber surrounding the membrane area to the outlet of the chamber.
The third measure is a thin protective sleeve with a cross-section slightly larger than that of the water strand. 
This sleeve is positioned around the water strand, effectively isolating it from the turbulence within the chamber. 
The final measure optimizes internal and external air pressure balancing. 
This is achieved by incorporating one check valve and two PTFE adhesive patches.
When combined, these mechanisms greatly reduce the impact of water accumulation on the strength of water streams. 
We detail the design of the chamber unit and present a set of quantitative experiments and a perception study to optimize the design parameters. 
The current JetUnit prototype can achieve a range of 16 to \SI{442}{\kilo\pascal} on-skin contact pressure and a maximum frequency of \SI{10}{FPS}. 

Additionally, we conducted a user study to investigate the ability of the JetUnit prototype to render various haptic patterns within a single VR story. 
Participants reported their experiences, particularly noting the degree of reality and enjoyment achieved through the integration of various haptics with our system.

\section{Related Work}
\subsection{On-body Force Feedback in VR}

Providing force feedback that matches the magnitude and duration of an interaction is key to enhancing realism in VR. 
Researchers have focused on developing haptic devices to deliver precise force feedback tailored to specific interactions, including both soft and gentle touches (e.g., \cite{Lee2015, 6548410, 7989976, Yoshida2019}) as well as intense impacts (e.g., \cite{9517129, tsai2022impactvest, tsai2019elasticvr, tsai2019elastimpact}). 
Since rendering gentle and intense haptics often requires distinct force activation mechanisms, much research addresses these separately, focusing on one group of force feedback at a time. 

Several studies have explored the rendering of gentle and soft touches to one's fingers. For example, TapeTouch \cite{zhu2022tapetouch} proposes using a piece of soft tape and varying its contour deformation to provide soft sensations upon touch. Suga et al. \cite{suga2023softness} demonstrate softness rendering to a finger by combining electro-tactile stimulation and force feedback. Similarly, Sonar et al. \cite{sonar2021soft} attach a thin piece of soft pneumatic actuator to one's fingertips, offering subtle sensations and tactile perception.

Other research aims to offer strong, often sudden forces for an immersive VR experience. For example, ImpactVest \cite{tsai2022impactvest}, ElasticVR \cite{tsai2019elasticvr}, and ElastImpact \cite{tsai2019elastimpact} render multilevel impact force feedback on the body, hands, or head, simulating experiences such as being shot, punched, or slashed by using independently controlled impactor blocks equipped with elastic bands. Motor-driven devices have also been widely used as a means to render strong forces. ExoInterfaces \cite{tsetserukou2010exointerfaces} and GuideBand \cite{tsai2021guideband}, for example, use DC motors in opposite directions on the upper arm to pull the user's forearm, generating strong forces. 
Recently, propeller thrust has garnered interest in generating strong force feedback. For example, Thor’s Hammer \cite{heo2018thor}, LevioPole \cite{sasaki2018leviopole}, AirCharge \cite{10.1145/3586183.3606768}, Wind-Blaster \cite{je2018wind}, and Aero-Plane \cite{je2019aero} attach varying numbers of propellers or air-jet compressors to handheld devices or directly to a user's wrist to provide strong force feedback in VR.

The last group of research focuses on rendering force with gradual changes. For example, Force Jacket \cite{delazio2018force} uses an array of pneumatically-actuated airbags to compress the users' body and arms, rendering force feedback with continuously changing levels. Similarly, Kanjanapas et al. \cite{kanjanapas2019design} render gradual changes in shear using 2-DoF pneumatic actuators.

As discussed earlier, while the aforementioned research covers a wide range of force intensities, frequencies of occurrence, and applications to different areas of the body collectively, few can encompass multiple types of force feedback within a single device.

\subsection{Mechanisms to Generate Force Feedback}
Given the diverse types of force feedback required in VR, researchers have proposed numerous mechanisms to generate them (e.g., \cite{kim2018hapcube, fang2021retargeted, choi2018claw, hosono2022feedback, Trinitatova2019, Camardella2022, 10160996}). One common method to simulate haptics is through various types of exoskeletons to provide controllable force feedback, augmenting the user's body displacement \cite{7922602}.
For example, HapticGEAR \cite{hirose2001hapticgear}, Naviarm \cite{maekawa2019naviarm}, and SPIDAR-W \cite{nagai2015wearable} all feature exoskeletons mounted on the back of the user. 
Other designs, such as CLAW \cite{choi2018claw}, RML Glove \cite{6750032}, and the work by Jo et al. \cite{7139982}, are mounted on the dorsal side of the user's hand. 
There are also exoskeleton haptic devices designed to be mounted on the user's finger, such as those developed by Perez et al. \cite{Perez2015} and Leonardis \cite{7177743, 7784835}.
Alternatively, through diverse types of linkage designs, exoskeletons can also be passive \cite{yukawa2010continuously, li2020haplinkage}, rendering force feedback without external power sources. Examples of such works include DigituSync \cite{nishida2022digitusync}, which shares hand poses between two users through its passive exoskeleton, inTouch \cite{brave1997intouch}, enabling passive interactions between users separated by distance, and HandMorph \cite{nishida2020handmorph}, designed to render haptic feedback from a smaller hand through a passive exoskeleton.
Recently, supplementary mechanisms, such as active brakes, have also been proposed as a means to enhance the capabilities of passive exoskeletons or to broaden the range of force feedback they can provide \cite{fang2020wireality, gu2016dexmo, hinchet2018dextres}.

Another popular method involves using electrical muscle stimulation (EMS) to directly stimulate the muscles with electrical pulses, providing perceived force feedback.
Farbiz et al. \cite{farbiz2007electrical} develop an EMS system that places electrodes on the arm to simulate the sensation of hitting a tennis ball.
Hosono et al. \cite{hosono2022feedback} use EMS to control muscle contractions for sharing tactile experiences on the fingertips.
Kurita et al. \cite{kurita2016stiffness} and Lopes et al. \cite{lopes2017providing} also use EMS to control muscle contractions, with Kurita focusing on creating the sensation of an object's stiffness and Lopes enabling users to feel the resistance and weight of virtual objects and walls.
EMS requires only an array of thin electrodes placed around the muscles \cite{pfeiffer2015cruise, tamaki2011possessedhand, takahashi2021increasing}; thus, compared to exoskeletons, its compact form factor makes it suitable for wearable haptic devices. However, as the electrical impulses travel through the skin, EMS may cause uncomfortable tingling sensations \cite{Pfeiffer2017, omura1985electrical} to some users.

The last group of mechanisms to highlight involves pneumatic.
Pneumatic haptic devices utilize dynamic air pressure to produce versatile forces or tactile sensations.
By precisely controlling the pressurization and depressurization of airbags, various studies have explored delivering different force feedback to the user’s head, wrist, arm, or torso.
For example, PneumoVolley \cite{gunther2020pneumovolley} delivers tactile sensations by varying air pressure around the head, allowing users to feel compressive forces and pressure changes.
Devices like PneuHaptic \cite{he2015pneuhaptic}, Siloseam \cite{10.1145/3357236.3395473}, Bellowband \cite{Young2019}, and Squeezeback \cite{10.1145/3025453.3025526} use compressed air to inflate or deflate pneumatic actuators placed around the wrist or forearm, creating localized pressure and vibration stimuli.
In addition to tactile sensations, pneumatic systems have demonstrated the capability to create strong force feedback, such as simulating rigid collision effects by combining additional embedded MR-brakes \cite{CinqMars2017}.
The Force Jacket \cite{delazio2018force} further develops a pneumatically actuated jacket for immersive haptic experiences, capable of rendering force feedback not only gentle interactions, like a hug, but also strong interactions, such as a snowball hitting the chest.
Despite their promising potential, pneumatic systems face significant drawbacks, including slow response times that limit their ability to render instant intense impacts and the complexity of integrating bulky compressors, which require substantial energy to pressurize the air to the desired level.

\subsection{Using Liquids in HCI for Interactions and Haptics}
Although not very common, liquids have been used in the field of HCI to facilitate interactions and render haptic sensations. 
One approach involves using water to facilitate interactions within water. For example, GroundFlow \cite{han2023groundflow} provides multiple-flow feedback by having users in VR actually step into a water-based haptic floor system. Similarly, Sinnott et al. \cite{sinnott2019underwater} propose an underwater VR system where users are immersed in water for buoyancy training. Combined with visual projections, the AquaTop display \cite{koike2013aquatop} enables users to interact with a visual-haptic interface by poking, stroking, or hitting the water's surface while taking a bath. Scoopirit \cite{matsuura2018scoopirit} allows users to scoop up water beneath a projected image, which becomes a mid-air image when raised.

Water has also been used to render tactile sensations on the body. Leveraging the sensations rendered by the flow of liquid, HydroRing and HapBead \cite{han2018hydroring, han2020hapbead} utilize the motion of liquid or small beads within microfluidic channels to deliver sensations as the liquid travels through. Other works explore sensations through dynamic changes enabled by liquid. For example, Chemical haptics \cite{lu2021chemical} proposes to deliver different types of liquid stimulants to the user's skin to render haptic sensations. ThermoCaress \cite{liu2021thermocaress} reproduces the illusion of a moving thermal sensation by moving the pressure stimulation with water. Therminator \cite{Sebastian2020} provides on-body thermal feedback through mixing the heated or cooled liquids. The capability to move mass has also been utilized to render dynamic changes in weight. For example, GravityCup \cite{cheng2018gravitycup} introduces a liquid-based handheld device that simulates realistic object weights and inertia shifts. Similarly, PumpVR \cite{kalus2023pumpvr} renders changes in weight by varying the mass of the controllers according to the properties of virtual objects or bodies.

In our work, we utilize water as the medium to provide force feedback for VR applications. Distinct from much of the existing research, our approach employs water jets---a stream of fluid projected through a nozzle that can travel distances and deliver varying forces without dissipating. Although water jets have been used in various applications, from fabrication \cite {natarajan2020abrasive} to massaging \cite{viitasalo1995warm, brandau1968performance}, their potential as wearable force feedback devices remains underexplored. Our work aims to fill this gap.

\section{JetUnit System Overview}\label{overview}

\begin{figure}[h]
  \centering
  \includegraphics[width=\linewidth]{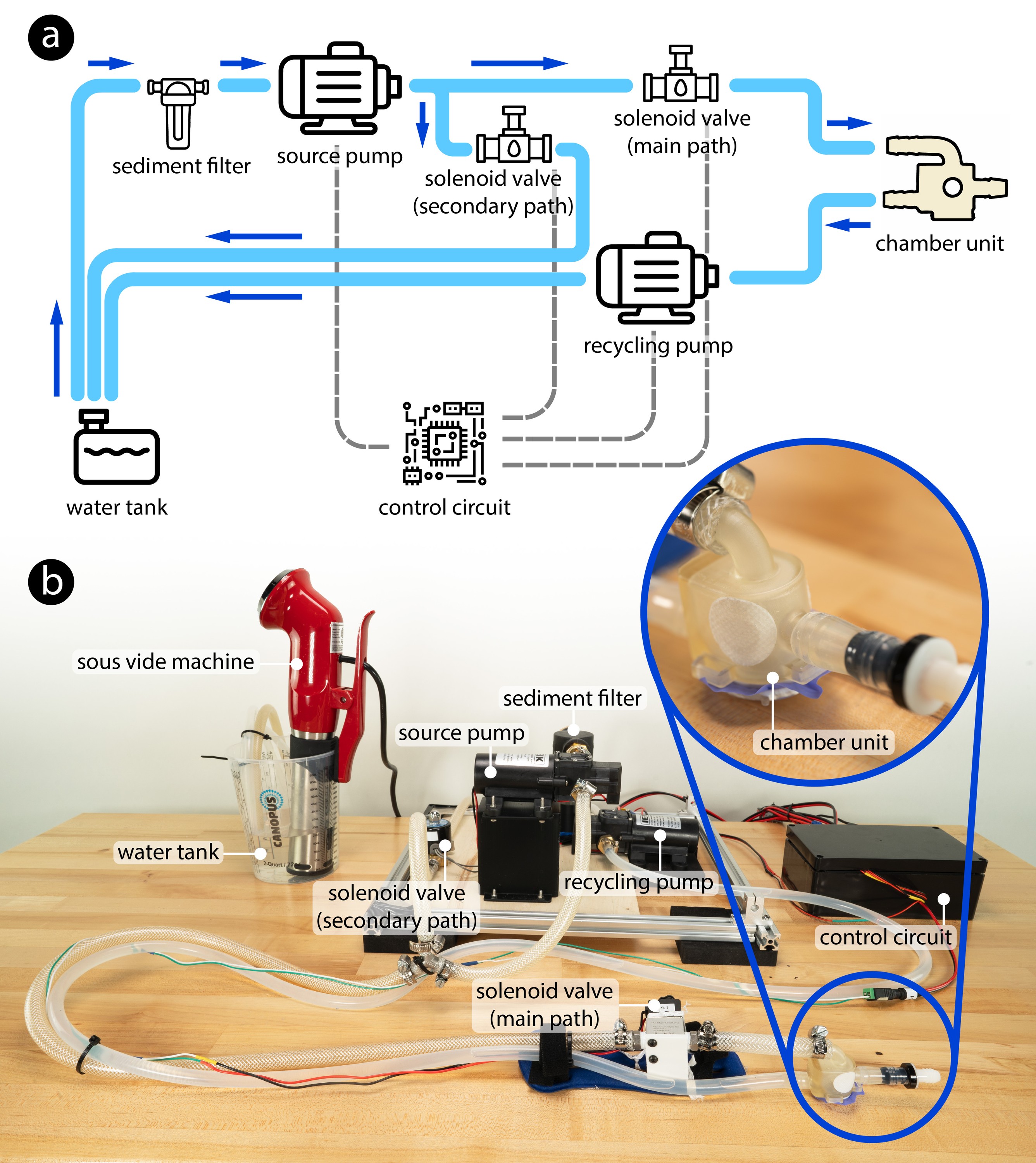}
  \caption{(a) Schematic of JetUnit System. (b) The JetUnit system setup.}
  \Description{Figure 2 illustrates (a) the schematic of the JetUnit system, and (b) the system setup}
  \label{fig:system_schematic}
\end{figure}

Figure \ref{fig:system_schematic}a illustrates the JetUnit schematic. 
It comprises four major components: a water source with a tank and water pumps, which circulate water throughout the system; a chamber unit that propels the water and provides force feedback to the user; a network of tubing that carries water to the chamber and recycles it back to the water tank; and a control circuit system. We briefly introduce each of the major components in this section. In Section~\ref{chamber unit design}, we detail the chamber unit design.

Given that the current JetUnit prototype is designed primarily for one haptic actuator, the entire system, including six meters of tubing, needs only \SI{0.6}{\liter} of water, with \SI{0.2}{\liter} stored in the water tank. 
The water tank is equipped with a sous vide machine to keep the water at a constant room temperature.
A \SI{116}{PSI} diaphragm water pump (IEIK) is used as the source pump to pressurize the water. It can produce a high flow rate, reaching up to 3.4 liters per minute in the 3/8 inch tubing. 
Between the water tank and the source pump, an additional sediment filter (LOVHO) is added to safeguard the source pump by filtering out debris carried by the water from the tank.

The water comes out of the source pump, flows through the network of tubing, and divides into two separate streams: the main path to the chamber unit that generates force feedback, and a secondary path that directly returns to the water tank. 
A pair of solenoid valves (Tailonz Pneumatic 2V025-08) controls these two paths. 
When the main path is open, the water enters the chamber unit, providing force feedback to the user, and then, through a recycling pump, returns to the water tank. 
The recycling pump, identical to the source pump, is specifically used to enhance the system's water recycling efficiency.

When no force feedback is needed, the main path is closed, allowing water to directly return to the water tank via the secondary path. 
This path ensures that at any given moment, water within the circulation does not accumulate and add unexpected pressure. 
It is important to note that when both paths are open, water can enter both paths simultaneously, resulting in a significantly reduced force exerted on the user. 
This is suitable for rendering soft and gentle force feedback. 
Detailed analyses of the force variations associated with these configurations are presented in Section \ref{sec:design_guideline}.

The water pumps and solenoid valves are controlled by an ESP32 microcontroller. 
A RoboClaw 2$\times$60A motor controller is used to regulate the speed of the source pump.
We note that the use of water pumps in the system inevitably introduces noise. 
To reduce this noise, both the source pump and the recycling pump are enclosed in a box lined with noise-canceling egg crate foam (WVOVW).
This box is supported by six shock-absorbing anti-vibration pads (MyLifeUNIT) to further mitigate vibrations from the pumps.

\section{Chamber Unit Design}\label{chamber unit design}
The custom-designed chamber unit is the key to rendering force feedback. 
It serves two main functions: providing force feedback with varying intensities and frequencies through water jets, while also keeping the user dry.

\begin{figure}[h]
  \centering
  \includegraphics[width=\linewidth]{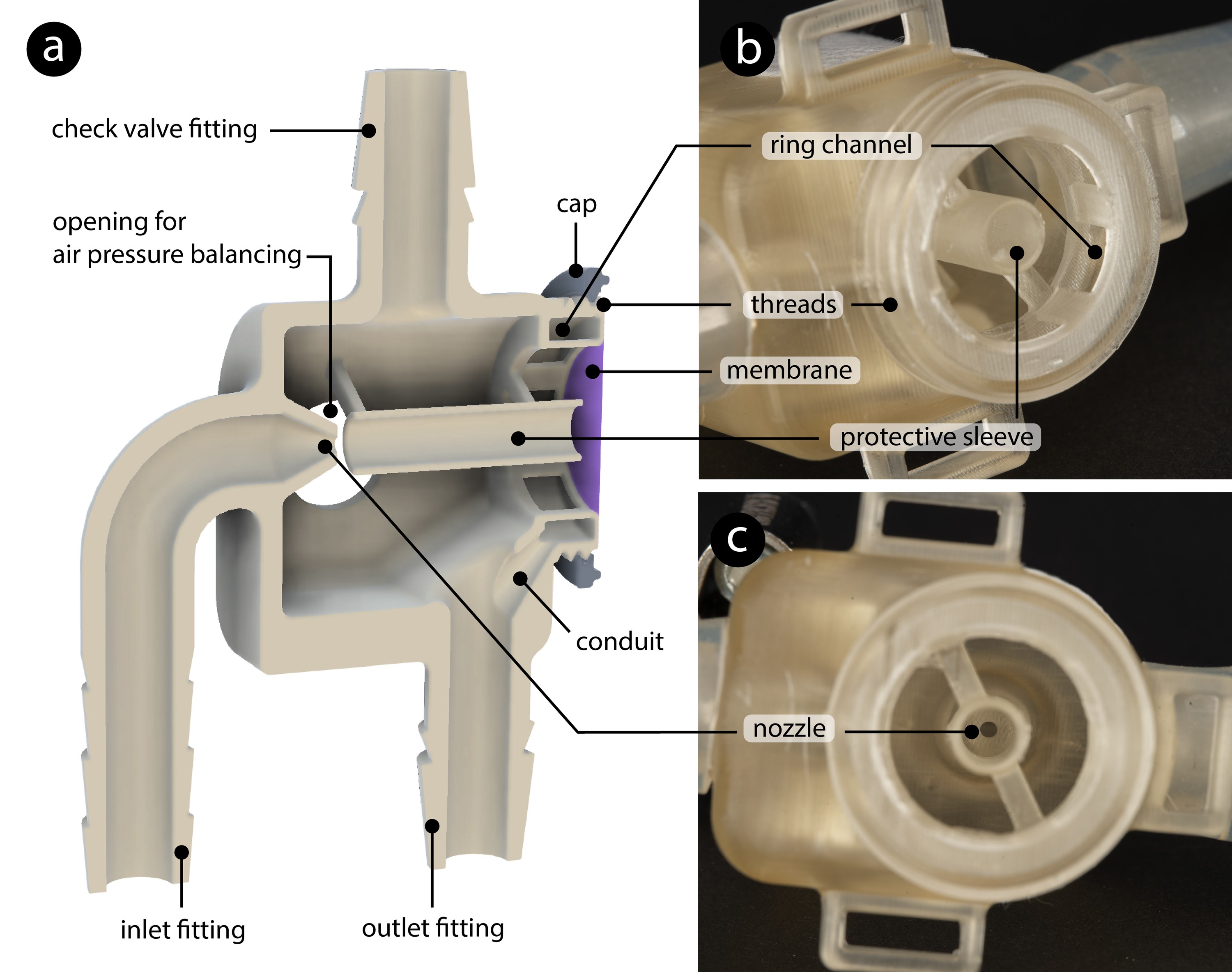}
  \caption{(a) Perspective view of a rendering of a half chamber unit cut from the middle. JetUnit chamber's views: (b) perspective, and (c) front view.}
  \Description{Figure 3 shows (a) a perspective view of a rendering of a half chamber unit cut from the middle. JetUnit chamber's views: (b) perspective, and (c) front view.}
  \label{fig:chamber}
\end{figure}

Figure \ref{fig:chamber} shows the chamber unit design. 
The chamber unit is a 25 mm $\times$ 28 mm $\times$ 32 mm quasi-cylindrical container with an opening at the front. 
A piece of elastic membrane is installed at the front to contain the water. 
A \SI{1.2}{\milli\meter} diameter nozzle is positioned at the back of the chamber, \SI{25}{\milli\meter} from the front opening, facing its center, and designed to emit water jets. 
A protection sleeve with a diameter of \SI{4.4}{\milli\meter} and a length of \SI{22}{\milli\meter} is positioned directly in front of the nozzle. 
The pressurized water jet from the nozzle enters the sleeve before coming into contact with any possible residual water inside the chamber. 
Additionally, a ring channel around the chamber's front opening is designed to direct water, deflected from the membrane in any direction, to the conduit that connects to the water outlet.
Finally, a check valve and two hydrophobic PTFE adhesive patches are installed to equalize the air pressure between the chamber and the atmosphere. 
The protection sleeve, air balancing features, and ring channel with the conduit to the chamber outlet collectively ensure that the water jet maintains its momentum before hitting the membrane. 
In addition, connecting the outlet of the chamber to the recycling pump further supports this momentum.
In the following subsections, we discuss the chamber design considerations in detail.

\subsection{Membrane Material and the Resulting Reduction in Force}\label{membrane}
The membrane at the front opening of the chamber unit plays a crucial role in retaining water while also delivering the water's impact to the user's skin. 
Thus, the material should be able to withstand high water pressure without breaking, while minimizing energy absorption from the material itself.

\begin{figure}[h]
  \centering
  \includegraphics[width=\linewidth]{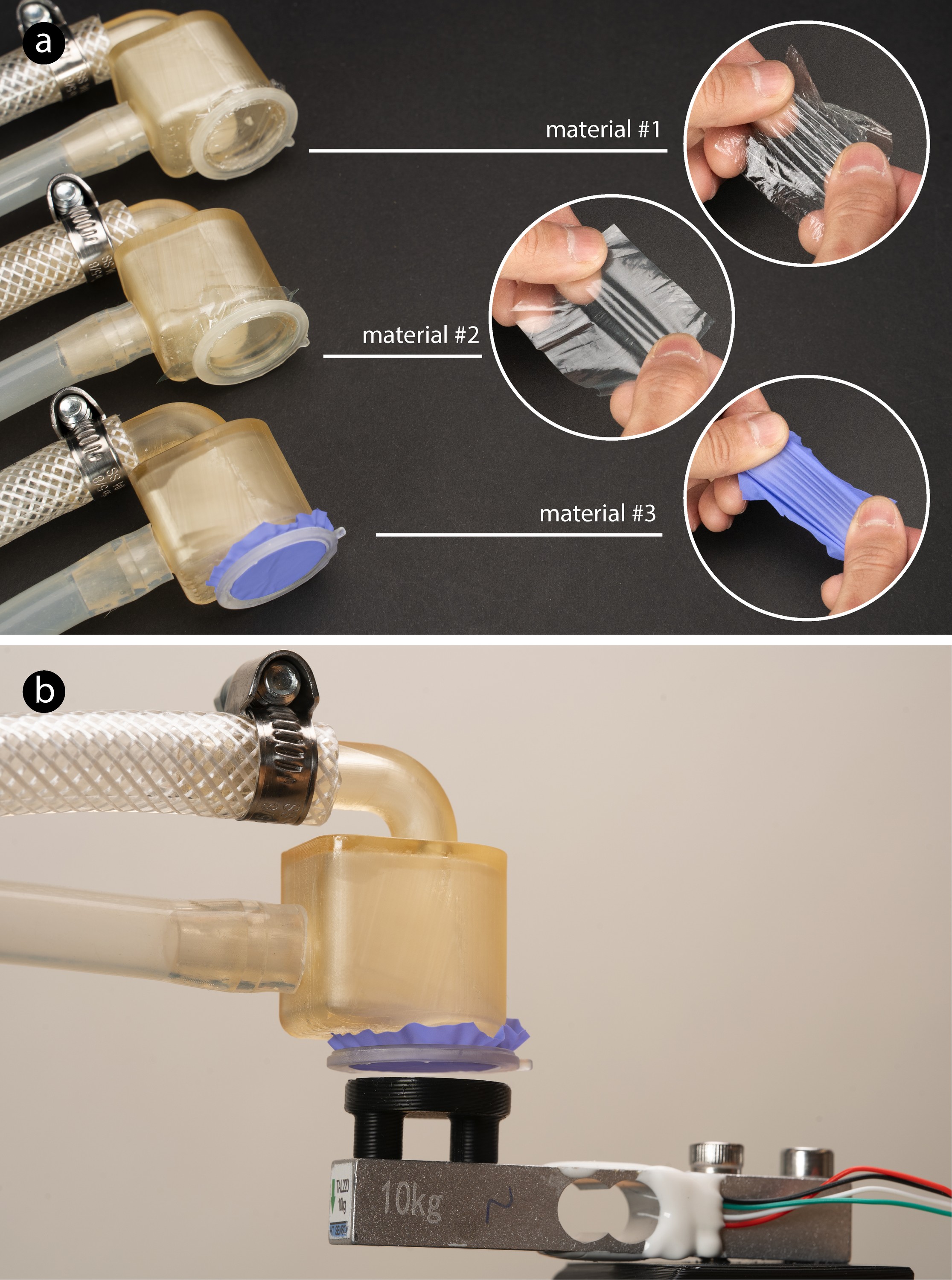}
  \caption{(a) Selection of membrane materials. (b) Setup for force measurement}
  \Description{Figure 4 shows (a) the selection of membrane materials. (b) Setup for force measurement.}
  \label{fig:loadcell}
\end{figure}

We considered three membrane materials that vary in thickness and elasticity, including ultra-thin, \SI{30}{\micro\metre} non-elastic low-density polyethylene (material \#1); \SI{150}{\micro\metre} non-elastic polyethylene (material \#2); and \SI{100}{\micro\metre} elastic nitrile butadiene rubber (material \#3), as shown in Figure \ref{fig:loadcell}a.
To decide on the material, we measured the force exerted after the water was jetted onto each membrane. 
We used the force of a bare water jet as a baseline (i.e., without a membrane).

\begin{figure}[h]
  \centering
  \includegraphics[width=\linewidth]{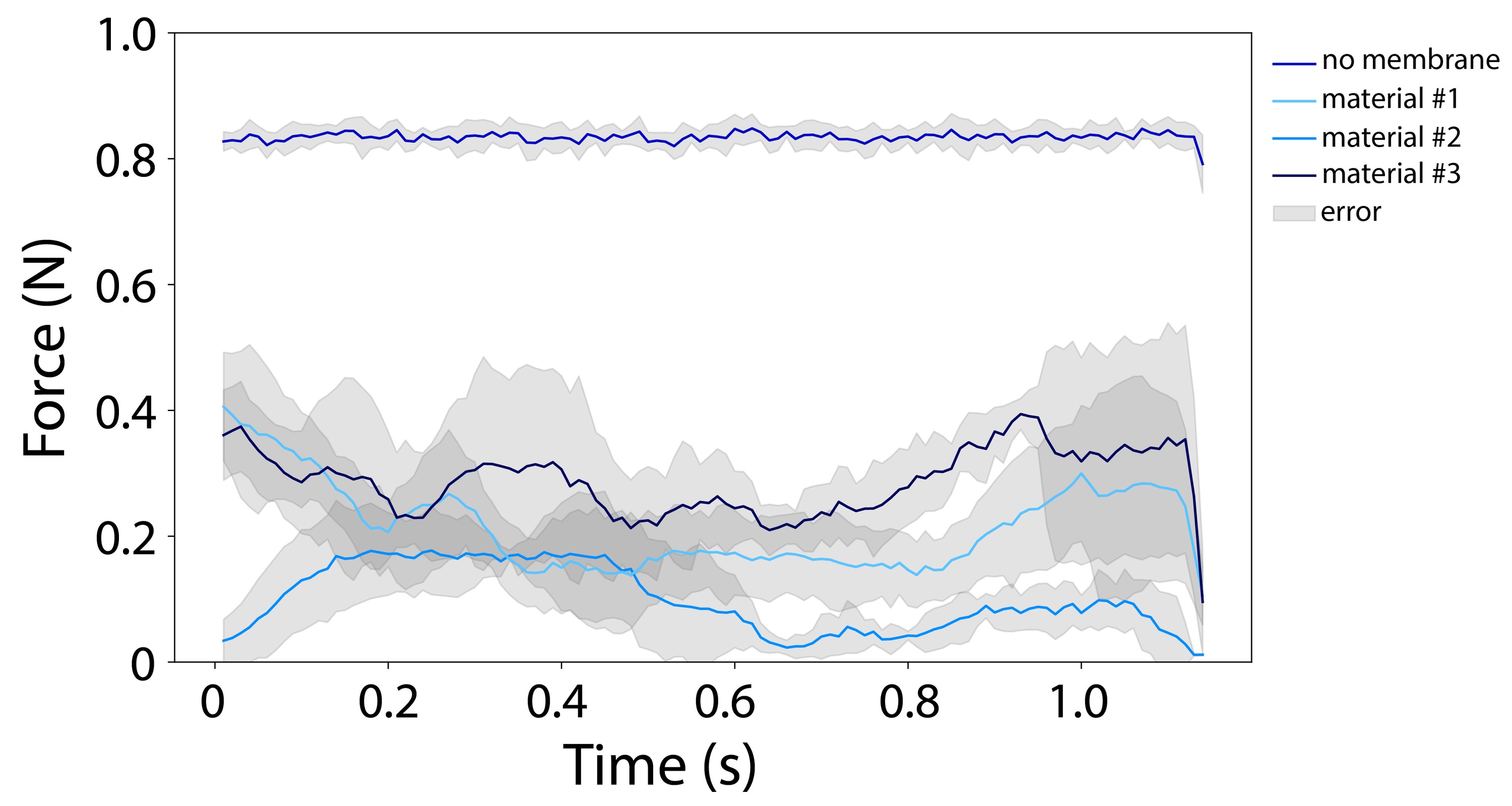}
  \caption{Force measurement results: The blue-tinted curves represent the force measurements for both the baseline and each membrane material; The grey-shaded area indicates the standard deviation of the force measurements at each timestamp.}
  \Description{Figure 5 presents force measurements for both the baseline and each membrane material. The baseline results are the highest, reaching approximately 0.84N. The results from all three membrane materials show lower magnitudes with large fluctuations.}
  \label{fig:membrane_measurement}
\end{figure}

Figure \ref{fig:loadcell}b illustrates the basic experimental setup for force measurement. 
A 3D-printed basic chamber unit (i.e., excluding measures for maintaining water jets' momentum)
was mounted horizontally against a straight bar load cell (SparkFun TAL220).
One side of the load cell was anchored at the edge of the table and the other side was hung freely. 
The load cell was calibrated with a gram-scale weight set, ranging from \SI{10}{\gram} to \SI{500}{\gram}. 
We activated the water jet five times, each for one second, during each round of measurements and calculated the average force reading.

The result is shown in Figure \ref{fig:membrane_measurement}. 
As expected, the baseline condition (i.e., bare water jets) presents the highest and most stable level of force among all conditions, at approximately \SI{0.84}{\newton}. 
All three types of membranes exhibit certain levels of force reduction, with materials \#1 and \#3 showing a force reduction of around 60\%, and material \#2 showing a reduction of around 87\%. 
Due to its thickness, the non-elastic material \#1 broke several times during our testing, leading us to choose the elastic nitrile butadiene rubber (material \#3) as the final membrane material.

\subsubsection{Issues arising from the use of the membrane}
While the chosen material \#3 outperformed the other candidates, using membrane has introduced several issues.

First, the force from the water stream was reduced by 60\% compared to the bare water jet baseline. 
Although the material itself would inevitably absorb some of the energy from the water impact, this significant reduction in force is largely attributed to water accumulating within the chamber, when the membrane is introduced. 
This accumulation acts as an additional buffer that dampens the water stream.

Second, our experiment revealed that force measurements under membrane conditions were unstable, exhibiting large fluctuations, as shown by the grey-shaded error area in Figure \ref{fig:membrane_measurement}. 
This instability was caused by water rebounding off the membrane, creating internal turbulence, and intermittently disrupting the flow of subsequent water jets from the nozzle.

Third, the use of membranes also introduced the potential for unbalanced air pressure within the chamber. 
For example, if the volume of water ejected from the nozzle exceeds the volume exiting the chamber, the increased volume of internal water causes a rise in air pressure, leading to the membrane bulging.
This bulging increases the area of contact with the skin. 
In contrast, if the outlet flow volume exceeds the inlet flow volume, the internal pressure decreases, causing the membrane to be sucked in and diminishing the force exerted.

In summary, a basic chamber unit sealed with the membrane but without additional measures would result in the force perceived by the user being significantly weaker and also very unstable. 
We will next introduce additional chamber designs to mitigate these issues.

\subsection{Minimizing the Impact of Accumulated Water within the Chamber Unit}
We implemented a total of four measures to reduce the impact of the accumulation of water within the chamber unit. 

\begin{figure}[t!]
  \centering
  \includegraphics[width=\linewidth]{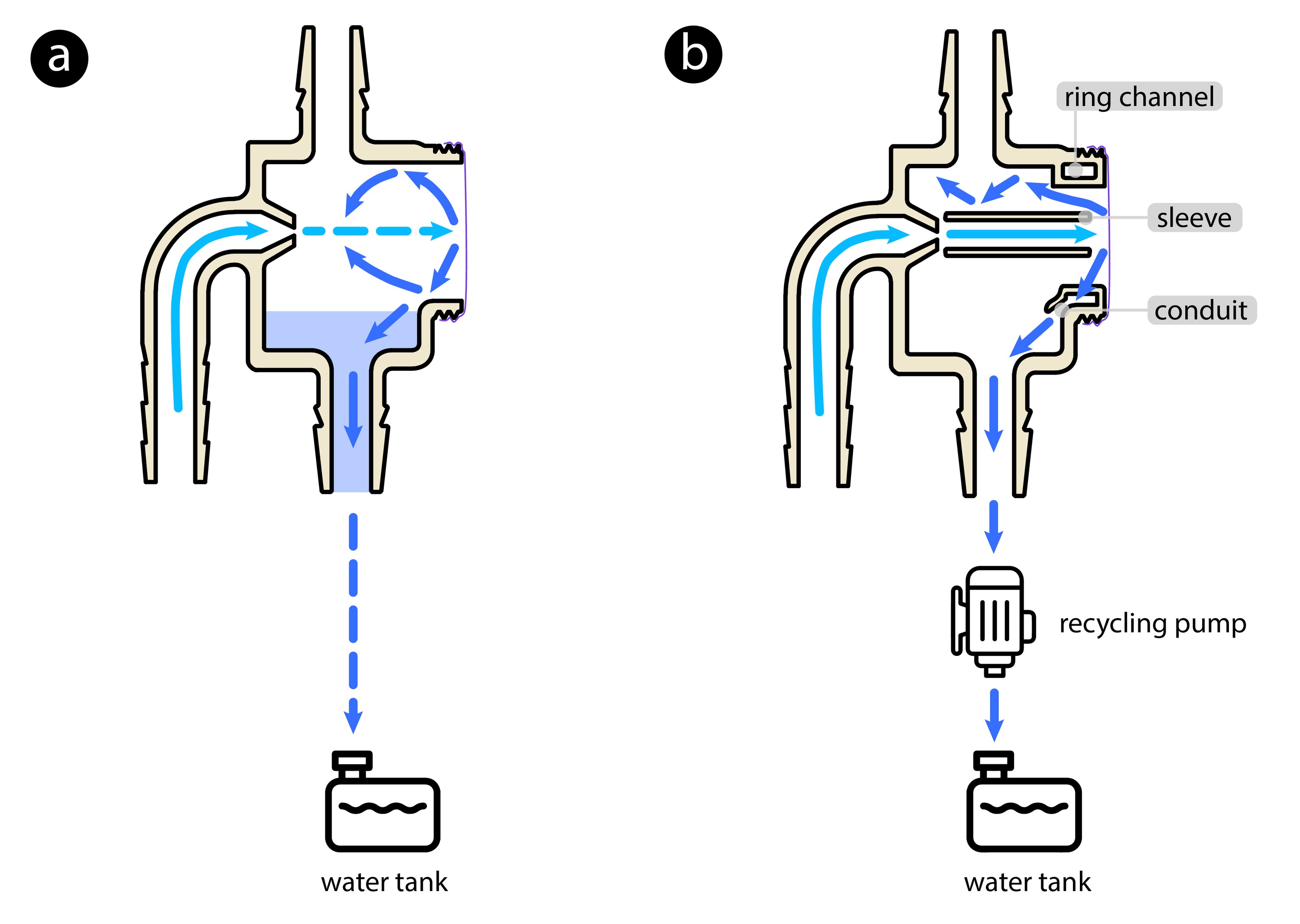}
  \caption{(a) Water accumulation, along with internal turbulence, is observed when the chamber opening is sealed with a membrane while the chamber outlet is directly connected to the water tank. (b) Connecting chamber outlet to a recycling pump, along with incorporating measures such as a protective sleeve, ring channel, and conduit, can enhance the efficiency of water egress.}
  \Description{Figure 6 illustrates (a) water accumulation, along with internal turbulence, is observed when the chamber opening is sealed with a membrane while the chamber outlet is directly connected to the water tank. And (b) connecting chamber outlet to a recycling pump, along with incorporating measures such as a protective sleeve, ring channel, and conduit, can enhance the efficiency of water egress.}
  \label{fig:egress_feature}
\end{figure}

First, as already introduced in Section~\ref{overview}, we added a recycling pump at the outlet of the chamber unit to remove accumulated water within the chamber as quickly as possible. 
Figure \ref{fig:egress_feature}a and b show a comparison of water accumulation with and without the recycling pump. 
Using the same basic chamber unit, when the source pump is activated at full power, the chamber unit without the recycling pump is filled with water within 0.6\,s. 
However, with the addition of the recycling pump, the chamber is never filled up.

Second, we designed a ring channel around the chamber's front opening to further accelerate the removal of residual water. 
Although the recycling pump can significantly reduce the accumulated water, some water still remains within the chamber. 
It is important to note that the remaining water cannot be easily removed with a more powerful recycling pump, due to the inherent design limitations of the chamber unit, where the water outlet cannot be positioned directly adjacent to the membrane. 
Thus, for certain angles of the chamber, such as when the membrane faces downward, water will inevitably accumulate until it reaches the height of the outlet before it can be suctioned out.

To address this, a ring channel is designed around the chamber's front opening area, as illustrated in Figure \ref{fig:chamber} and Figure \ref{fig:egress_feature}b. 
This ring channel features six openings, each measuring \SI{4}{\milli\meter} in height and \SI{5}{\milli\meter} in width, evenly distributed around the inner sidewall. 
When water is stranded near the membrane, this ring channel allows water to escape through it, before further accumulation happens.

Third, to further prevent the turbulence caused by the rebounded water from interfering with the incoming jet flow from the nozzle, we designed a protective sleeve between the nozzle and the chamber front.
The sleeve, shaped like a hollow cylinder, has an inner diameter of \SI{4.4}{\milli\meter}---slightly larger than the nozzle---with a wall thickness of \SI{0.6}{\milli\meter} and a length of \SI{22}{\milli\meter}. 
One end of the sleeve is situated just \SI{0.6}{\milli\meter} away from the chamber front.
When water is pumped out of the nozzle, the sleeve effectively isolates the water stream from any potential accumulated water within the chamber.

Finally, to optimally balance the external and internal air pressure of the chamber unit while activating the recycling pump, we incorporated two circular openings of \SI{10}{\milli\meter} diameter into the side wall of the chamber unit.
These openings are covered with biomedical scientific hydrophobic polytetrafluoroethylene (PTFE) adhesive patches.
These patches, which boast a filtration rate of 99.97\%, are capable of filtering particles as small as \(0.3 \, \mu\text{m}\), thus helping to maintain pressure equilibrium while also preventing water leakage through the openings.
In addition, we installed a one-way check valve (B08FJ1TSSJ) on the chamber unit to improve air flow into the chamber, particularly during periods of negative pressure inside the chamber caused by activating the recycling pump.

\subsubsection{Improvement}
After implementing all the upgrades to the chamber configuration, we compared force measurement results under three conditions: one with bare water jets from a \SI{1.2}{\milli\meter} nozzle placed \SI{25}{\milli\meter} away from the load cell; another with the same nozzle inside a chamber \SI{25}{\milli\meter} away from the thin elastic membrane; and a third condition featuring the same chamber configuration as the second, but additionally outfitted with a recycling pump and the upgraded configuration (i.e. the sleeve, the ring channel and conduit, and the pressure-balancing opening). 
We activated the water jet five times at the full capacity of the source pump for one second in each of the three conditions and calculated the average value of the force readings for comparison. 

\begin{figure}[h]
  \centering
  \includegraphics[width=\linewidth]{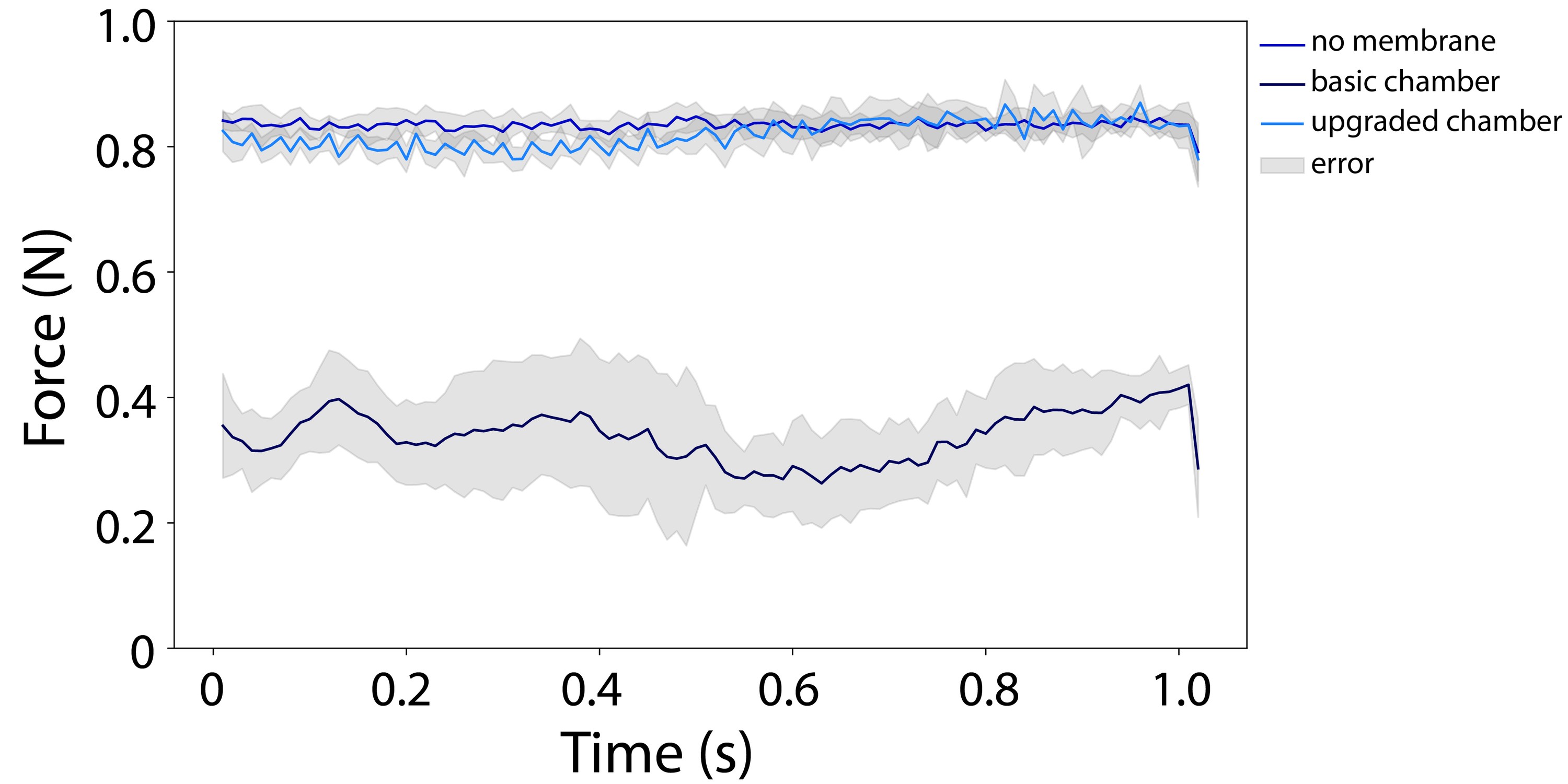}
  \caption{Force measurement results: The blue-tinted curves show the force measurements for the baseline, the basic chamber condition, and the upgraded chamber condition. The grey-shaded area represents the standard deviation of the force measurements at each timestamp.}
  \Description{Figure 7, the blue-tinted curves show the force measurements for the baseline, the basic chamber condition, and the upgraded chamber condition. The grey-shaded area represents the standard deviation of the force measurements at each timestamp.}
  \label{fig:refinedMeasurePlot}
\end{figure}

The results are shown in Figure \ref{fig:refinedMeasurePlot}. 
The average force impact of the bare water jets is \SI{0.84}{\newton}. 
The basic chamber, which lacks efficient water egress measures, could only produce an average force impact of approximately \SI{0.33}{\newton} from the water jets. 
However, the upgraded chamber can now provide an average force impact of about \SI{0.82}{\newton}, which is nearly identical to the effect of bare water jets.

\subsection{Choosing the Nozzle Dimensions}
\subsubsection{Equation of continuity}\label{subsec:nozzle_dimensions}

In fluid dynamics, the conservation of mass principle dictates that mass is conserved within a control volume for constant-density fluids. 
That being said, in a given water source pump system, the mass flow rate at the nozzle opening remains constant. 
Therefore, when the nozzle diameter changes, the flow rate of the water jets changes inversely to maintain this constant mass flow rate. 
Typically, a smaller nozzle diameter will result in higher water pressure and velocity, leading to greater impact force at the point of contact. 
However, it is also important to balance this with the capabilities of the water source pump. 
When the nozzle is too small, it can overload the pump system by significantly increasing the resistance to flow, resulting in a decrease in water pressure and velocity. 
Thus, selecting the optimal nozzle diameter is a balance between the pump's characteristics and the desired jet force.

\subsubsection{Measuring the force of different nozzle sizes}

\begin{figure}[h]
  \centering
  \includegraphics[width=\linewidth]{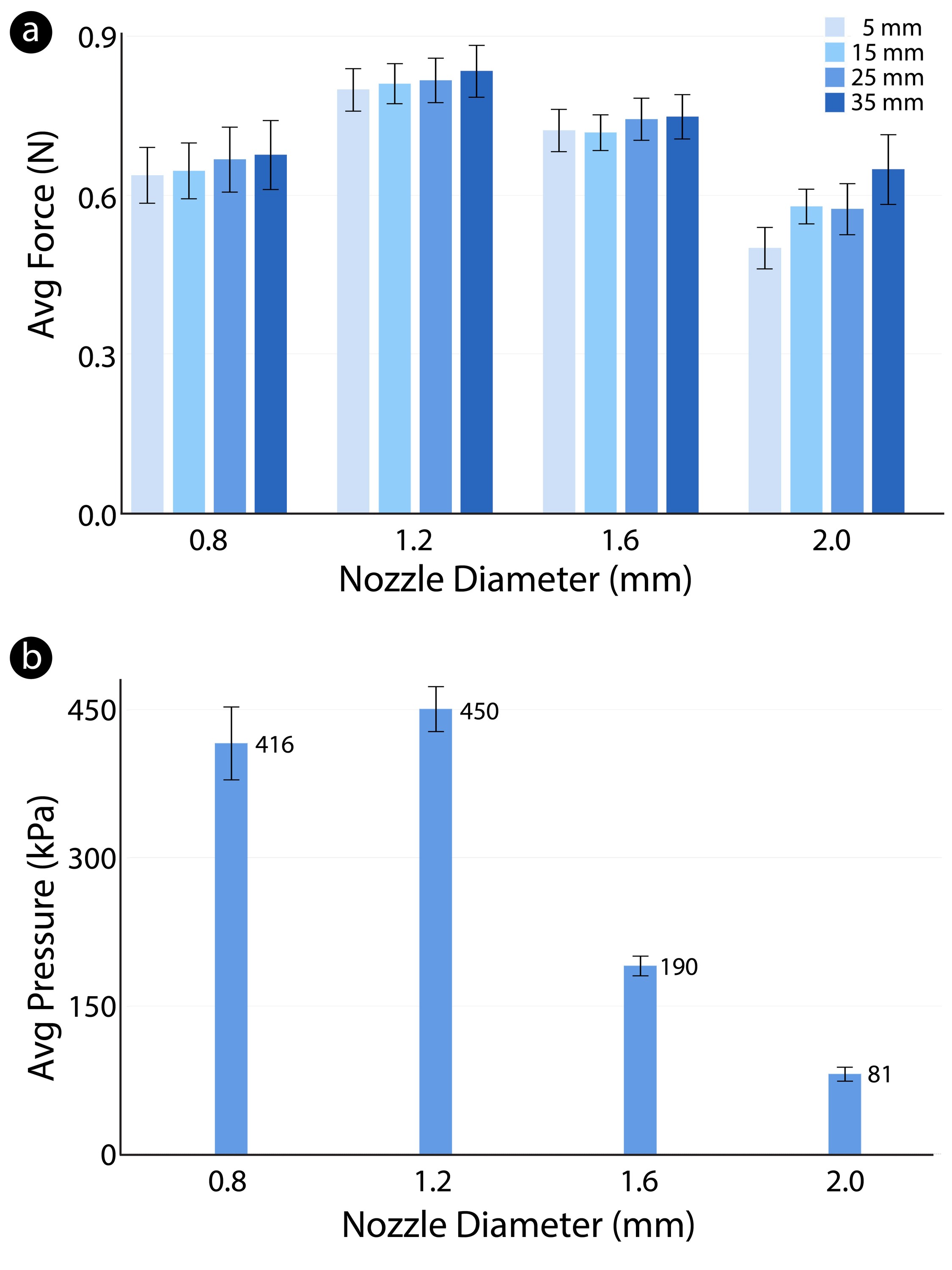}
  \caption{Measurements of (a) the average force magnitude and standard deviation at different nozzle-to-contact area distances with respect to nozzle diameter; (b) the average pressure and standard deviation at a fixed 25 mm nozzle-to-contact area distance with respect to nozzle diameter.}
  \Description{Figure 8 presents the measurement results of (a) the average force magnitude and standard deviation at different nozzle-to-contact area distances with respect to nozzle diameter. (b) the average pressure and standard deviation at a fixed \SI{25}{\milli\meter} nozzle-to-contact area distance with respect to nozzle diameter.}
  \label{fig:nozzle_diameter}
\end{figure}
\begin{figure}[h]
  \centering
  \includegraphics[width=\linewidth]{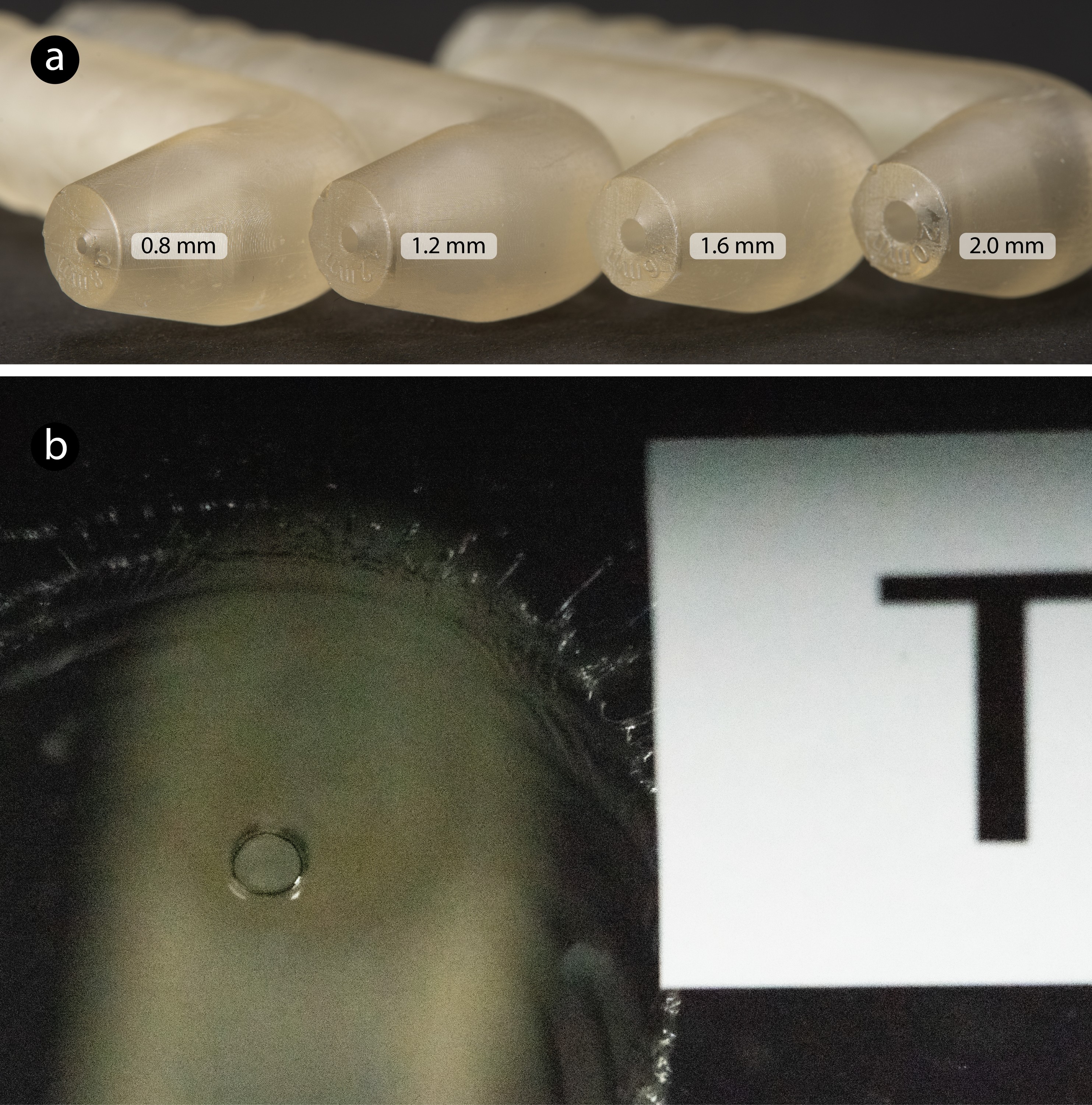}
  \caption{(a) A set of nozzle prototypes, ranging in size from 0.8 mm to 2.0 mm, is used to measure the actual contact area of water jets. (b) The set of nozzle prototypes was positioned 25 mm away from a transparent acrylic board, onto which a letter `T' was affixed for calibration. The horizontal line of the letter measured 4.7 mm in length, and the vertical line was 4.9 mm long. Water jets were then activated, and photographs were taken. The diameter of the actual contact area can be determined through pixel-to-actual length conversion.}
  \Description{Figure 9 (a) shows the set of nozzle prototypes used for measuring the actual contact area. Figure 9 (b) is a sample photo used for determining the actual contact area.}
  \label{fig:contact_area_setup}
\end{figure}

To optimize the design of the nozzle dimensions, we tested four nozzles with diameters of \SI{0.8}{\milli\meter}, \SI{1.2}{\milli\meter}, \SI{1.6}{\milli\meter}, and \SI{2.0}{\milli\meter}, as illustrated in Figure \ref{fig:contact_area_setup}a. 
Considering that the distance between the nozzle and the membrane surface might also influence the water jet's velocity upon contact, we assessed the force impact of water jets from each nozzle at distances of \SI{5}{\milli\meter}, \SI{15}{\milli\meter}, \SI{25}{\milli\meter}, and \SI{35}{\milli\meter}. 
For each condition, We recorded the real-time impact forces of two-second water jets five times using the setup detailed in Section \ref{membrane}.

As shown in Figure \ref{fig:nozzle_diameter}a, with the same nozzle diameter, the nozzle-to-contact area distance has a relatively modest influence on the average force of water jets. 
To balance the arrangement of tubing fitting barbs on the chamber with its compactness, we ultimately chose a nozzle-to-contact area distance of \SI{25}{\milli\meter}. 
This distance provides adequate space for the necessary configuration without affecting the force of the water jets that much.

Decreasing the nozzle diameter from \SI{2.0}{\milli\meter} to \SI{1.2}{\milli\meter}, as shown in Figure \ref{fig:nozzle_diameter}a, results in an increase in the average force of the water jets at the same nozzle-to-contact area distance, following the fluid dynamic principles discussed earlier. However, when the nozzle diameter is further reduced to 0.8 mm, the force exerted by the water jets decreases, indicating an overburden on the pump system. 

\subsubsection{Measuring the pressure}
It is important to note that the absolute force measurement is not the only correlation with the perception of the force on the skin. 
Considering the small contact area of the water jets, we decide to use pressure as a more relevant metric to assess human perception \cite{1287224}. 
Furthermore, to ensure that the user does not experience discomfort or pain due to the small contact area, it is important to maintain the pressure generated by our system below the pain-pressure threshold (PPT) \cite{fransson1993sensitivity}.
As our water jet flow rate is high and the water's travel distance is short (25 mm), if we neglect air resistance and spreading, then the estimated contact area diameter should be close to the nozzle dimension. 
However, the actual contact area will be slightly larger due to the spread of the water stream, influenced by factors such as air resistance, surface tension, and the breakdown of the stream into droplets. 
For an accurate pressure estimation, we measured the actual diameter of the contact area, as shown in Figure \ref{fig:contact_area_setup}b. Specifically, with the water flow from the nozzle size ranging from 0.8 to \SI{2.0}{\milli\meter}, the corresponding contact areas measured were \SI{1.4}{\milli\meter}, \SI{1.5}{\milli\meter}, \SI{2.2}{\milli\meter}, and \SI{3.0}{\milli\meter}, respectively.

Pressure is defined as the force per unit area, using the formula
\[ P = \frac{F}{A} \]
In our case, the measured delivered force is divided by the effective contact area to estimate the pressure.
Figure \ref{fig:nozzle_diameter}b shows the average pressure estimation for different nozzle diameters. 
The results indicate that the highest average pressure, 450~\text{kPa}, was generated using a nozzle diameter of \SI{1.2}{\milli\meter} positioned \SI{25}{\milli\meter} away from the membrane. The second highest pressure, 416~\text{kPa}, was observed with a nozzle diameter of \SI{0.8}{\milli\meter}. The pressure produced by nozzles with diameters of \SI{1.6}{\milli\meter} and \SI{2.0}{\milli\meter} was much lower than that of the previous two conditions.

\subsubsection{Perception study}
\label{subsec:perception_study}
Furthermore, we conducted a basic perception study to collect participant feedback to better understand the varied force sensations associated with different nozzle diameters. This user study was approved by the Institutional Review Board (IRB) of our university.

\begin{figure}[h]
  \centering
  \includegraphics[width=\linewidth]{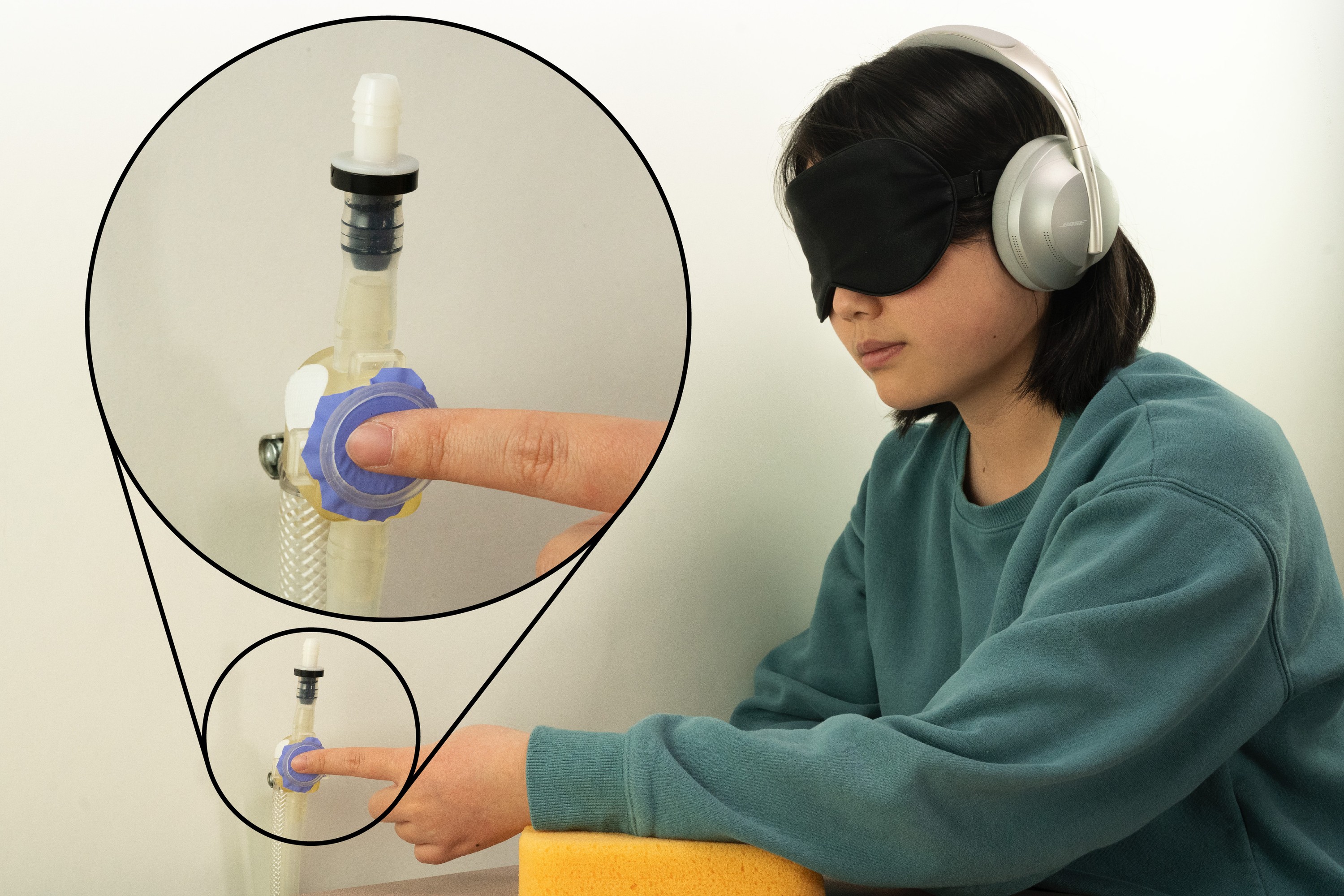}
  \caption{The basic perception study setup with zoom-in view at chamber location.}
  \Description{Figure 10 shows the basic perception study setup with a zoom-in view at the chamber location.}
  \label{fig:perception_study_setup}
\end{figure}

\textit{Participants}:
Participants (N = 7; 5 females, 2 males), aged 23-31 (Mean = 26.71, SD = 2.75), were recruited for this study and compensated at a rate of 15 dollars per hour. Three participants had experience with vibrational haptics, and the remaining four participants had varied experiences in haptics. All of them use their right hand as their dominant hand.

\textit{Procedure}: 
Participants were instructed to wear an adjustable eye mask to block their vision and noise-canceling headphones, which played light music, to minimize influence from the surrounding environment.
Participants were instructed to rest their dominant arms on the table with palms facing sideways while the JetUnit was placed on the user's index finger.
We chose the fingertip as the force feedback testing area because it is known to be one of the most sensitive areas of the human body with a low PPT \cite{fransson1993sensitivity}.

According to our pressure estimations shown in Figure \ref{fig:nozzle_diameter}b, the highest average pressure produced by our system is within a safe zone, ensuring that it does not harm the user. 
The study involved testing with three chamber units in total, each equipped with a nozzle of different diameter as mentioned above: \SI{0.8}{\milli\meter}, \SI{1.0}{\milli\meter}, and \SI{1.2}{\milli\meter}. 
The source pump of the system was operated at full capacity to produce the maximum force impact of the water jets. 
The chambers were tested in a randomized order and each underwent five repetitions, resulting in a total of 15 trials per participant.

After each trial, participants were asked to assess the perceived intensity of the force at their fingertip using a free magnitude scale \cite{delazio2018force}, which allows a more natural and subjective evaluation of haptic perception. 
They were also asked to rate their comfort level on a Likert scale ranging from -3 to 3 to indicate their comfort level. 
After completion of all trials, the participants were interviewed for more detailed feedback. The entire study lasted approximately 40 minutes.

\textit{Results and discussion}:
Since participants use their own scales, we normalize these diverse ratings to a common scale for comparison using the Z-score normalization method. The normalized rating of perceived force intensity, \( Z \), is given by
\[ Z = \frac{X - \mu}{\sigma} \]
where \( X \) is the rating reported by participants, \( \mu \) is the mean of all ratings, and \( \sigma \) is the standard deviation of all ratings. 

\begin{figure}[h]
  \centering
  \includegraphics[width=\linewidth]{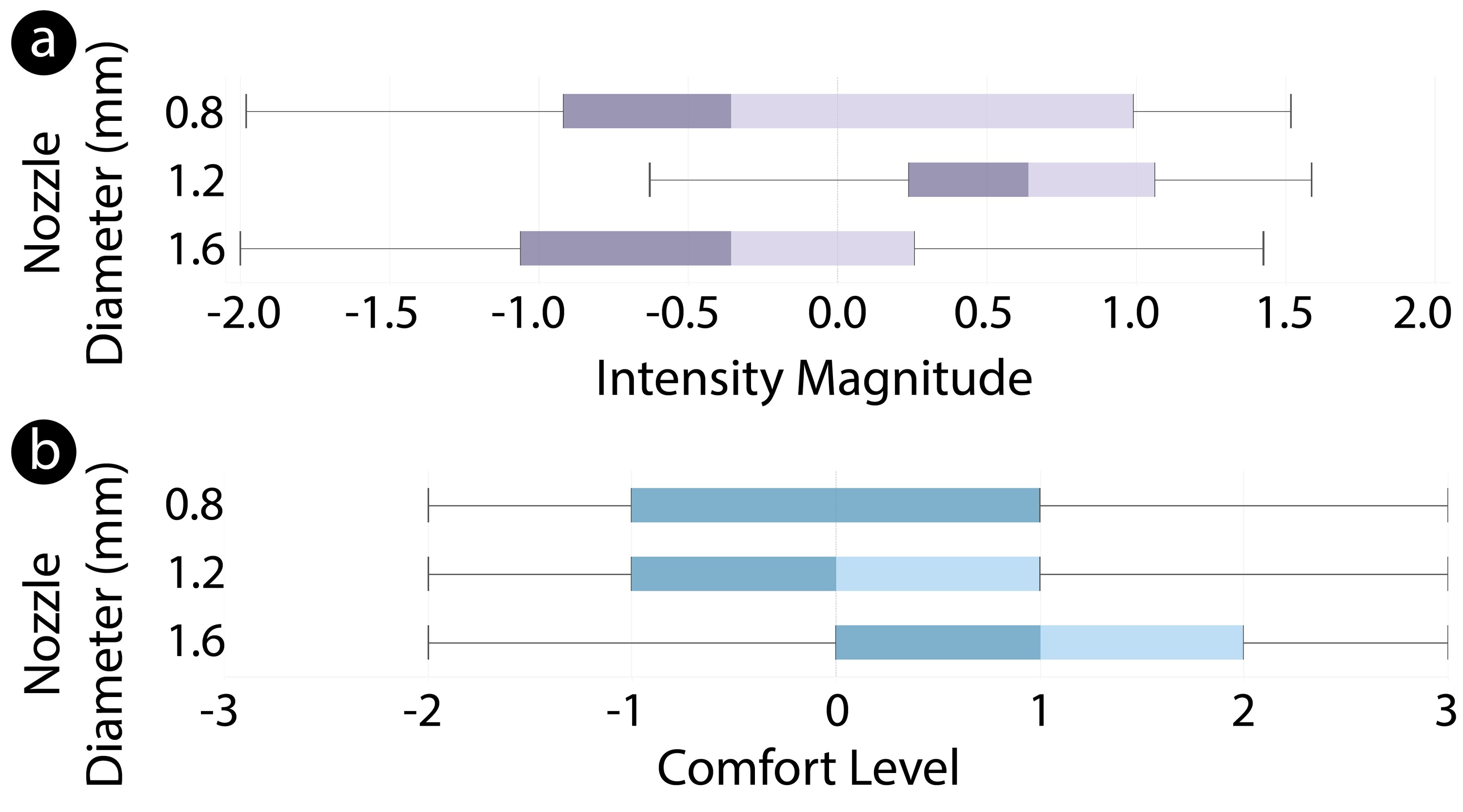}
  \caption{Boxplots represent (a) the normalized perceived force intensities rating and (b) the comfort level distribution.}
  \Description{Figure 11 shows the results from the basic perception study: (a) a box plot shows the normalized perceived force intensities rating and (b) a box plot shows the comfort level distribution.}
  \label{fig:perception_result}
\end{figure}

As shown in Figure \ref{fig:perception_result}a, after applying Z-score normalization, the ratings for all nozzle diameters fell within a range of -2 to 2. 
Among these, the nozzle diameter of \SI{1.2}{\milli\meter} showed the most notable perceived force intensity ratings. Given that the perceived force intensity ratings did not satisfy the normality assumption, as indicated by the Shapiro-Wilk test results ({\slshape W = 0.93, p < .05}), we ran a nonparametric analysis. 
The Friedman test revealed significant differences in perception among the use of three nozzle diameters ({\slshape p < 0.001}). Further analysis using Wilcoxon Signed-Rank Tests with Holm-Bonferroni adjustments for pairwise comparisons showed significant differences in perception between the nozzle diameters of \SI{1.2}{\milli\meter} and \SI{0.8}{\milli\meter} ({\slshape p < 0.05}), and between the nozzle diameters
of \SI{1.2}{\milli\meter} and \SI{1.6}{\milli\meter} ({\slshape p < 0.005}). However, the difference in perception between the nozzle diameters of \SI{0.8}{\milli\meter} and \SI{1.6}{\milli\meter} was not statistically significant ({\slshape p > 0.05}).

This perception result matches the calculated pressure estimate presented in Section \ref{subsec:nozzle_dimensions}, demonstrating that with the chosen pump in our system, setting the nozzle diameter to \SI{1.2}{\milli\meter} can produce the maximum force impact and pressure sensation.

The comfort level ratings corresponded to the perceived force intensities on their fingertips: a higher perceived force intensities rating corresponded to a lower comfort level (Figure \ref{fig:perception_result}). Despite participants assigning negative values to their comfort levels, during the interview session, they clarified that this primarily pertained to the effort needed to counteract the force that pushed their fingers away from the chamber membrane. All participants stated that the force and pressure exerted by the JetUnit did not cause any hurt or pain.

\subsection{Summary of Chamber Unit Design Optimization}
The chamber unit's design was optimized by implementing several key improvements. A recycling pump was added at the chamber outlet to quickly remove accumulated water, preventing internal turbulence and ensuring consistent force delivery. A ring channel around the membrane further expedited water egress, while a protective sleeve between the nozzle and the chamber front isolated the water jet from residual water. Pressure-balancing openings with hydrophobic PTFE patches and a one-way check valve were included to maintain air pressure equilibrium.

Measurements determined that a \SI{1.2}{\milli\meter} nozzle diameter positioned \SI{25}{\milli\meter} from the membrane offered the highest and most stable force impact (\SI{0.82}{\newton}), matching the bare water jet. A perception study confirmed that this setup provided the strongest force sensation without discomfort.

These enhancements resulted in a finalized chamber unit capable of producing robust and responsive force feedback. 
All hardware schematics and models are made available and open source \footnote{\url{https://github.com/znzhang26/JetUnit.git}}.

\section{Characteristics of the JetUnit}
\label{sec:design_guideline}

To quantify the spectrum of haptic patterns that can be generated by our JetUnit system, we measured them in terms of pressure and pulsing frequency.

\subsection{Range of Force Impact}

\begin{figure}[h]
  \centering
  \includegraphics[width=\linewidth]{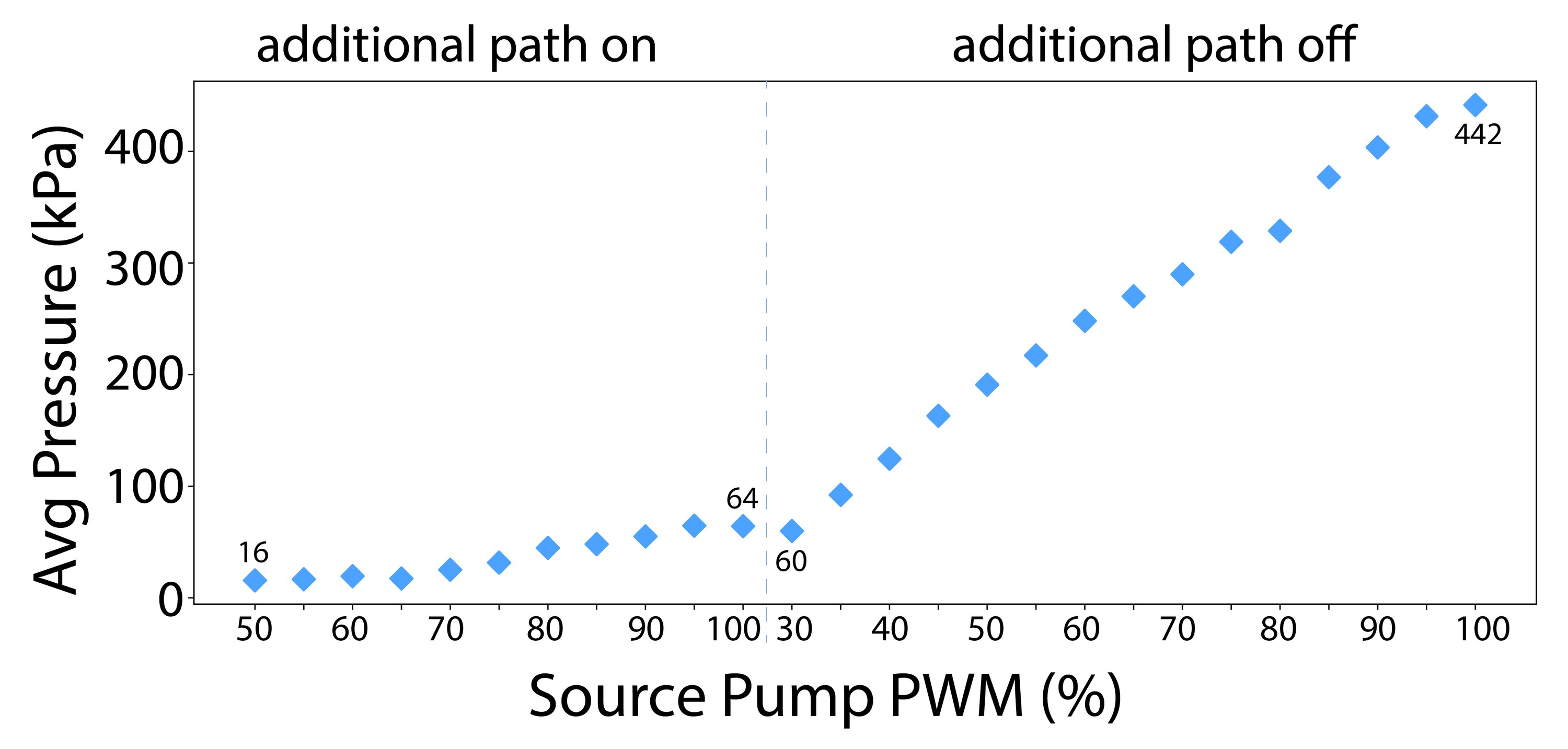}
  \caption{Full range of pressure produced by the JetUnit chamber.}
  \Description{Figure 12 shows the full range of pressure produced by the JetUnit chamber. The minimum pressure is 16 kPa and the maximum pressure is 442 kPa.}
  \label{fig:range_force}
\end{figure}

The JetUnit system is capable of providing a wide range of force feedback, measured in pressure. As shown in Figure \ref{fig:range_force}, the lowest average pressure achievable with our current implementation is 16~\text{kPa}. 
This level of pressure is achieved by opening the secondary path and setting the pulse width modulation (PWM) of the source pump to 50\%. 
Reducing the source pump's PWM further does not deliver sufficient water to form a water stream. 
In contrast, closing the secondary path significantly increases water flow to the main path, which in turn increases the pressure at the membrane. The pressure ranges from 60~\text{kPa} to 442~\text{kPa} when the PWM of the source pump is adjusted from 30\% to 100\%.

We should note that the maximum pressure that we can currently render is just below the average PPT in humans. 
Although our aim is to render a strong impact with water jets, we do not intend to cause any discomfort to the users. 
However, as the sensitivity of the haptics varies greatly among users, the maximum pressure can be easily increased, if needed, by switching to a more powerful water source pump. 
For comparison, consider the HB21000 Jacuzzi pump, which has a maximum flow rate of 246 liters per minute and is 72 times more powerful than our current source pump. Thus, the JetUnit is very capable of rendering stronger forces if required.

\subsection{Frequency of Pulsing}

Our JetUnit system can produce short, impactful bursts; continuous, long-lasting impacts; and repeated force feedback, such as high-frequency pulsing, by adjusting the switching frequency of the solenoid valve.

\begin{figure}[h]
  \centering
  \includegraphics[width=\linewidth]{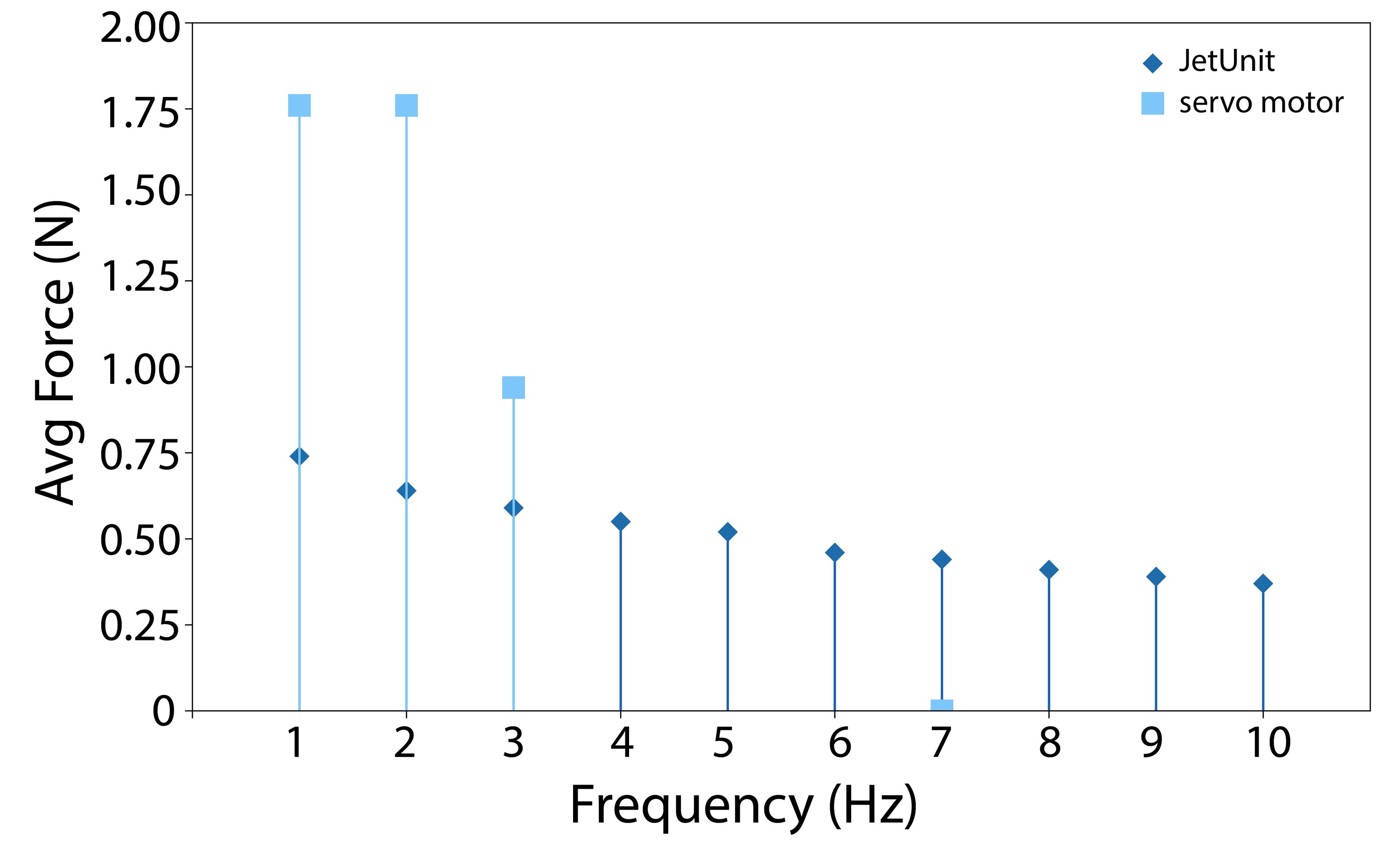}
  \caption{Comparison of JetUnit and servo motor in terms of achievable pulsing frequency and corresponding average force magnitudes at each frequency.}
  \Description{Figure 13 shows the results of the comparison of JetUnit and servo motor in terms of achievable pulsing frequency and corresponding average force magnitudes at each frequency.}
  \label{fig:range_pulsing}
\end{figure}

Specifically, JetUnit can reach pulsing frequencies of up to \SI{10}{Hz}, as shown in Figure \ref{fig:range_pulsing}. Although this frequency is not as high as that achievable with a vibration motor, it surpasses the pulsing sensations that can be rendered by common exoskeleton devices using servomotors. 
Figure \ref{fig:range_pulsing} presents a simple comparison of the pulsing frequency generated by a lightweight high-torque servo motor (DM90S), whose no-load working speed is \(0.10 \, \text{s} \, \text{per} \, 60^\circ\), with a stall torque of $2.0\,\text{kg}\cdot\text{cm}$. 
By setting its rotation angle range at \( 35^\circ \), we ensure that there is no contact with the target object during lifting. 
Although in theory this type of servo motor could achieve a maximum pulsing frequency of about \SI{7}{Hz}, in practice our measurement setup was unable to record any readings at this frequency. 
The actual highest recorded pulsing frequency was less than \SI{4}{Hz}, which is less than 40\% of the achievable frequency with our JetUnit prototype.

\subsection{Patterns with Gradually Changing Force}
\begin{figure}[h]
  \centering
  \includegraphics[width=\linewidth]{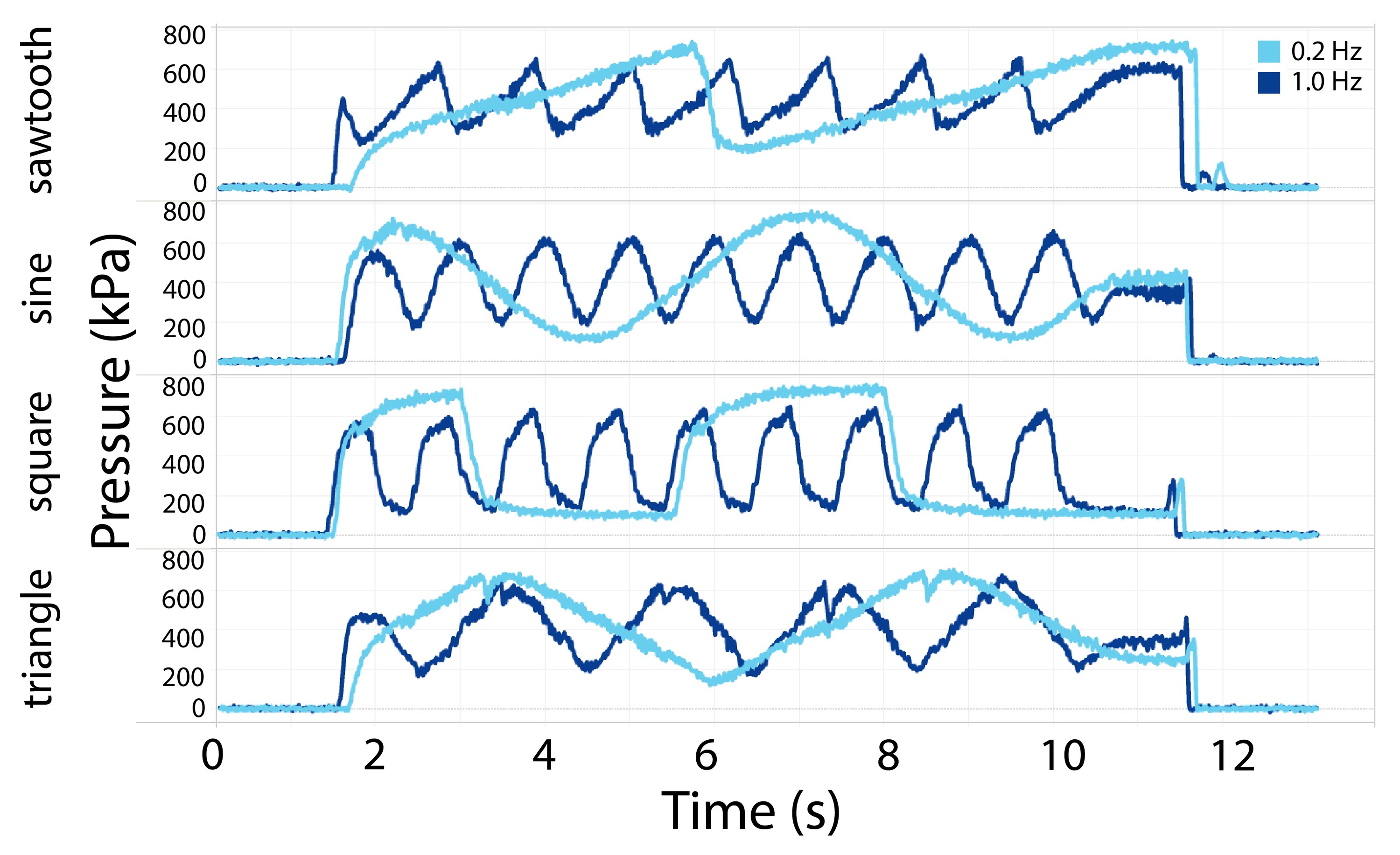}
  \caption{Real-time pressure measurements for haptic waveform patterns in sawtooth, sine, square, and triangle waves.}
  \Description{Figure 14 shows the real-time pressure measurements for haptic waveform patterns in sawtooth, sine, square, and triangle waves.}
  \label{fig:waves}
\end{figure}

With the ability to vary intensity and frequency, the JetUnit system can produce force patterns that gradually change, including linear, sine wave, triangle wave, square wave, and sawtooth wave patterns, across multiple cycle frequencies (Figure \ref{fig:waves}). 
When combined, this library of diverse force feedback types is well-suited for scenarios with complex interaction demands, such as accurately matching actions like touching, pressing, or even striking with sudden changes, on the dorsal side of the dominant hand.

\section{VR Experience Study}
We conducted a user study to investigate the JetUnit prototype's ability to render various haptic patterns. 
Specifically, we examined whether participants in VR could perceive different types of haptic feedback and if these varied force feedback sensations could provide enjoyment and a sense of realism. 
To achieve this, we developed a VR story that required different interactions from the user with their hand. 
Meanwhile, a single JetUnit device provided various haptic feedbacks on the dorsal side of the hand, matching the VR story.

Note that the decision to render the force feedback on the dorsal side of the hand was influenced by the VR scenario. Initially, we considered three different locations for the haptic rendering: the chest \cite{delazio2018force, tsai2022impactvest}, the forearm \cite{he2015pneuhaptic, lu2021chemical, Yoshida2019, 8816170}, and the dorsal side of the hand \cite{10.1145/2993148.2993171, jung2022wireless, 10.1145/3197768.3197785, 10.1145/1593105.1593195, MAZZONI20169}. In our preliminary exploration, the JetUnit device could render haptic feedback to all three locations. However, designing convincing VR scenarios for the forearm and chest proved challenging due to their larger skin areas compared to the small haptic rendering area provided by a single-chamber unit. 
In contrast, the dorsal side of the hand has previously been used in VR haptics and has a relatively small skin area, making it more suitable for our study with the single JetUnit implementation.

\begin{figure}[t!]
  \centering
  \includegraphics[width=\linewidth]{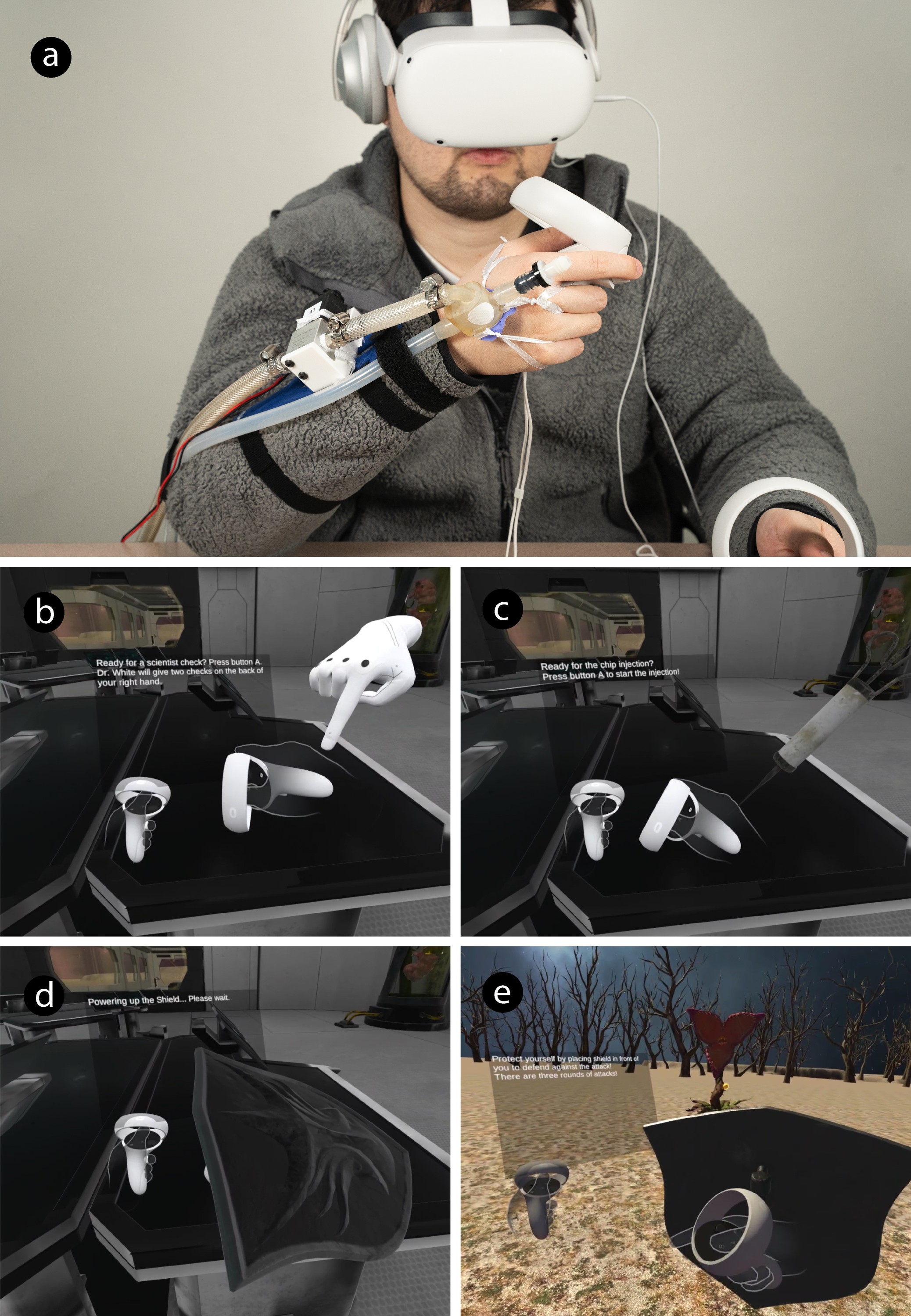}
  \caption{(a) VR user study setup. VR scenes: (b) user experiencing gentle touch, (c) user receiving needle injections, (d) user activating a power shield, and (e) user defending against flower-shaped monster attacks.}
  \Description{Figure 15(a) displays the setup for the VR user study, while panels (b) to (d) illustrate the tasks designed within the VR story.}
  \label{fig:vr_user_setup}
\end{figure}

\subsection{Participants}
Participants (N = 11; 8 females, 3 males), aged between 24-29 years (Mean = 26.91, SD = 1.76), were recruited for this study and compensated at a rate of \$15 per hour. All participants were right-handed, and none had any history of hand injuries. Among them, two had no previous VR experience, two had limited VR experience, and the rest had VR experience. 
Three participants experienced mild 3D motion sickness while using driving simulators, while the others reported no such issues.

\subsection{Procedure}

Participants began with a training session to familiarize themselves with the various haptic patterns they might encounter during the study.
Afterward, they were instructed to wear the chamber unit on the dorsal side of their dominant hand. 
The chamber unit was secured with adjustable bands and ribbons to ensure optimal fit and comfort. 
Participants were equipped with a VR headset and noise-canceling headphones before being guided to start the VR game. Upon completing the game, they were asked to answer both Likert scale and open-ended questions designed to collect feedback on their VR experience and thoughts on perceived haptics.

\subsection{Task}
The VR story developed for this study is structured around a single mission divided into three scenes. 
Participants, cast as `the chosen' warrior, are endowed with `magic' technology by a group of scientists to defeat a flower-shaped monster.

In the first scene, participants enter a laboratory where two scientists tap the backs of their hands twice in quick succession, over a short duration of 1.8\,s---one scientist applies gentle pressure, while the other applies intense pressure. Subsequently, participants receive a needle injection to implant `magic' enhancement fluid beneath the skin on the dorsal side of their hands. 
At the end of the first scene, that is, after injection, participants are prompted with a questionnaire to select the series of haptic sensations from the options provided that best match their experience.

In the second scene, participants are provided with three different shields to protect themselves from the flower-shaped monster. 
These shields differ in appearance. 
As participants explore the different shield options, a distinct set of haptic feedback patterns is activated: a sine wave, a square wave, or a sawtooth wave. 
After participants choose their shield, a questionnaire pops up asking them to identify the haptic pattern of each shield.

In the final scene, participants encounter three rounds of attacks from a flower-shaped monster, with pollen hitting the shield with different frequencies (2 Hz, 3 Hz, and 7 Hz). 
After surviving all three rounds, they report the number of different frequencies perceived during the three rounds of attack. 

\begin{figure}[b]
  \centering
  \includegraphics[width=\linewidth]{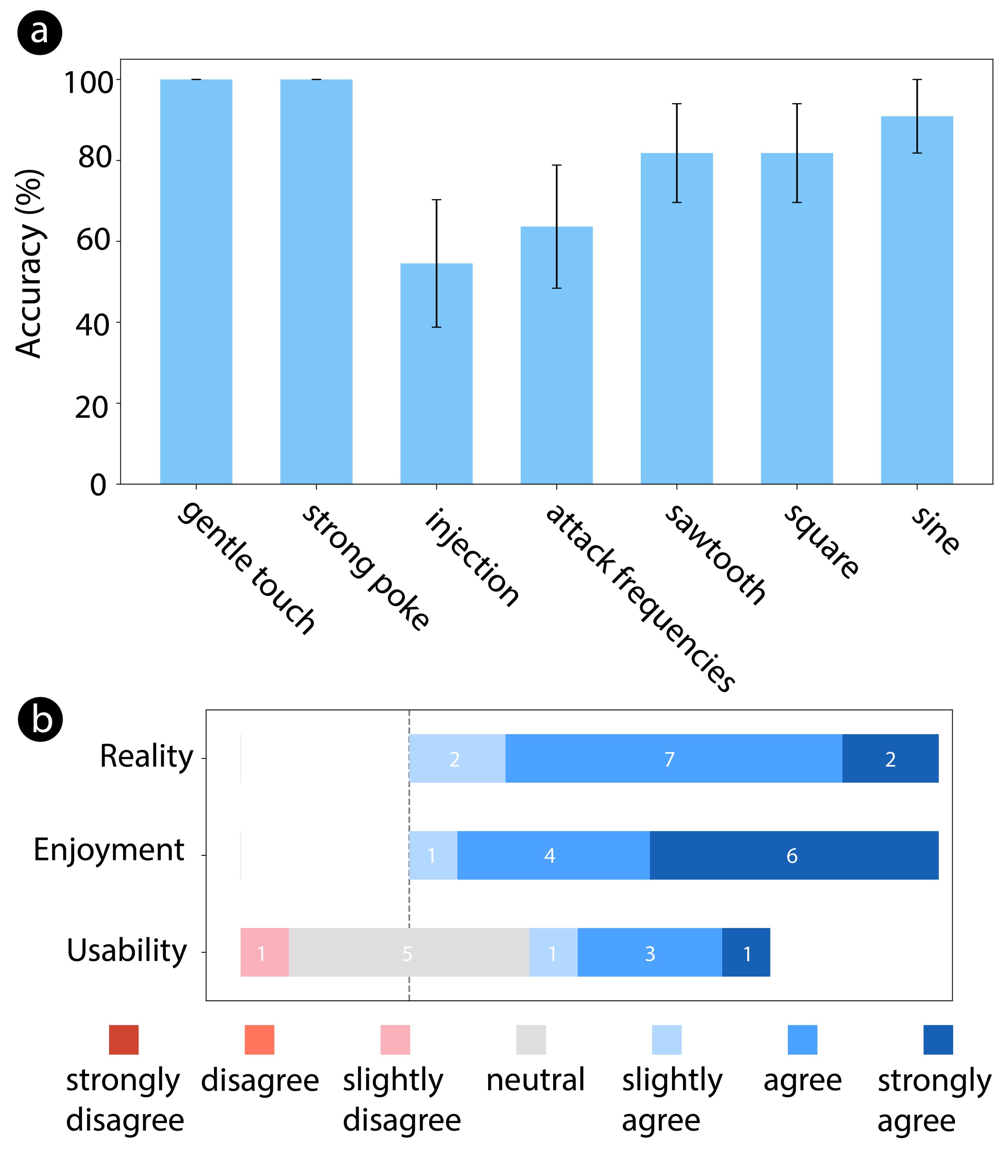}
  \caption{VR study results. (a) Perception accuracy. (b) Self-reported ratings.}
  \Description{Figure 16 shows the study results of (a) perception accuracy according to different haptic patterns, and (b) self-reported ratings according to VR haptic experience.}
  \label{fig:vr_results}
\end{figure}

\subsection{Results and Findings}
The results are presented in Figure \ref{fig:vr_results}. Figure \ref{fig:vr_results}a shows the accuracy of the participants' perception in response to various haptic patterns rendered to their dorsal side of the hand, which vary in perceived force intensities, frequency of occurrence, and waveform patterns. 
Short force durations (1.8\,s) for both gentle touch and strong poke are easily distinguished by users, achieving 100\% perception accuracy. 
Longer force durations, where water jets maintain contact with the participants' hands for 4.5\,s to simulate the sensation of liquid injection, revealed that 54.5\% of participants correctly reported gradually increasing contact pressure, while 45.5\% perceived the pressure as constant throughout the injection. 
This discrepancy might be attributed to constant visual rendering throughout the injection, which could send mixed signals to participants. 
Although participants did not perceive the change in pressure, they still reported enjoying it. 
The accuracy of perceiving waveform patterns is relatively high. Both sawtooth and square waveforms were reported with 81.8\% accuracy, and the sine waveform is the easiest to perceive among these three patterns, as its accuracy reaches 90.8\%. 
All participants were able to perceive the pulsing patterns, while 63.5\% successfully perceived all three frequencies. 27.3\% could only distinguish two frequencies, as the lower two frequencies were set similarly (2 Hz and 3 Hz), compared to the highest frequency defined in the game (7 Hz).

As shown in Figure \ref{fig:vr_results}b, all participants rated their VR experience as realistic. They mentioned that the haptic patterns matched their expectations during interactions in the virtual environment, enhancing their immersion in VR.
P1 mentioned, \textit{``The injection on my hand felt very realistic!''}
P2 said, \textit{``... When I felt the pollens shooting on my hand, actually, the frequency, is so sharp, so intense. It just felt very new, very novel, very realistic... The design of the game fits the nature of the hardware... I'm wearing the shield and getting shot and that makes everything organic and makes everything a good combination.''}
P6 detailed that \textit{``I felt realistic because with the visual that I was seeing and with the haptic feedback that I was getting, it kind of matched that what I would expect and when I would expect a touch to happen... Oh, the person is going to touch and then I actually feel the touch... And talking about the touch, say for example, the visual was this (the hand moves in a short distance for a gentle poke) and this (the hand moves in a longer distance for a stronger poke compared with the previous one) matched well with my expected touch feeling in terms of the strength.''}

All participants in the study enjoyed exploring the virtual environment, with more than half (54.5\%) rating their enjoyment as ``strongly agree.'' 
One participant (P3) was observed repeatedly switching the shields for four rounds, and emphasized her enjoyment in sensing the waves of haptic patterns and imagining the flow as it powered up - \textit{``I think the most interesting part is powering up the shield... I spend a lot of time experiencing the three types of shields... I selected the third shield and I really like the way it's being powered up and it even makes me feel I was powering up.''}
P6 said that \textit{``It was enjoyable because I have been using VR before, but now I have this additional haptic feedback on the body which adds like a new sensation. And it's not just one kind of feedback cause I felt pollen shooting, person touching, injection and having a shield, and feeling the hittings on the shield. They are all different perceptions, which were kind of a match to what I would expect if they were happening in real. So the enjoyment was this combination of multiple haptic feedback that I felt along with the visual.''}

In terms of device usability, participants' ratings are varied. P6 said that \textit{``The JetUnit device is easy to use just like you put on the VR headset. All you have to do is put on your hand and you are ready to go... And it syncs easily with the headset.''} 
The reason for deducting the easy usability is mainly the weight and flexibility concern. 
P7 reported that \textit{``The device itself has weight. 
It feels natural to use it as a shield. But at the same time, if you use it to simulate something lighter, it wouldn't make much sense because of the weight.''} 
P10 suggested making the future version portable as he felt that the pad, which fixes the solenoid valve to the arm, restricts movement to some extent.

Overall, the study confirmed that the JetUnit prototype can deliver a variety of haptic patterns, featuring a broad spectrum of forces and distinct pulsing frequencies. This versatility opens up opportunities to enhance user enjoyment and realism by enabling diverse haptic feedback across a range of interaction scenarios.

\section{Discussion}
\subsection{Limitations}
Although the haptic patterns provided by the JetUnit device are generally acknowledged, the system does have its limitations. 
One of the primary challenges is wearability, particularly due to the solenoid valve, which must be positioned as close to the chamber as possible to minimize latency caused by the travel time of the water flow in the tubing. This could be mitigated by using a lighter valve, although it would increase the cost of building the system.

The current setup complicates user mobility and overall comfort due to wire and tubing entanglement, which is further hampered by stationary water tank and pumps.
To address these issues, implementing the solenoid valve with a wireless control circuit could significantly reduce the risk of wire entanglement, thereby enhancing user movement flexibility. Additionally, considering that the system operates with a relatively small amount of water, there is a viable opportunity to integrate a portable water tank and pump directly onto the user's body, which also helps to reduce tubing entanglement.
Moreover, the current attachment mechanism could be improved for a better user experience. 
Replacement of the existing ribbon with an adjustable buckle strap could improve both ease of use and comfort, making the device more practical for extended use.

Another major limitation of the current setup is the selection of the placement of the device. 
As the key contribution of this work is the self-contained chamber design, we have focused solely on the single-unit implementation on the dorsal side of the dominant hand. 
However, our design is not limited to this placement. 
For rendering haptic feedback on larger skin areas like the forearm, chest, or even back, multiple chamber units arranged in arrays can be applied, which we will discuss in the following section.

\subsection{Future Directions}\label{future}
There are several promising directions for the development of the JetUnit system. 
Given the compact and small size of the chamber unit, it has the potential to design and deploy full-body haptic systems.
An approach to achieving a full-body haptic system with a single-unit implementation is to allow the chamber unit to move around the body \cite{10.1145/3550323, 10.1145/2984511.2984531}, reaching target areas as needed. 
Another solution is to create a multi-unit system. 
By creating an array of chambers that can be applied to various parts of the body \cite{delazio2018force, tsai2022impactvest}, we can simulate different environmental conditions, such as light and heavy rain, across large areas. 
This would involve multiple units working in concert, better enhancing the immersive experience.

As highlighted by Participant 5, \textit{{``I was hoping that the force feedback from JetUnit device is not only on a specific part of my hand, instead, my hand in general.''}} 
There is a desire for force feedback that is not limited to a specific part of the hand, but encompasses the entire hand.
Considering that sensation can vary significantly between different body parts, future iterations of the JetUnit system should feature adjustable maximum pressure and customizable strength settings, accommodating the varying haptic sensitivities of different body parts. This adaptability is important for achieving a more comprehensive and effective haptic experience.

Another area of future enhancement involves the integration of the temperature switching functionality \cite{Sebastian2020, liu2021thermocaress}. Leveraging water's ability to transmit temperature changes can significantly enhance the realism of virtual reality environments, adding a new dimension to the user experience. This functionality would allow users to feel temperature variations corresponding to different virtual scenarios, further immersing them in the environment.

Moreover, exploring ways to vary the contact area on the skin is another potential upgrade for the JetUnit system. By altering nozzle dimensions or configurations, the system could render a wider spectrum of haptic patterns. 
However, this would require addressing challenges related to water sealing and ensuring that the system remains reliable and effective despite structural modifications.

\section{Conclusion}

We investigated a haptic solution that utilizes a water system, designed to create a range of diverse haptic patterns. 
This system achieves a broad spectrum of perceived force intensities and pulsing frequencies of haptic rendering. 
It also enables the rendering of force feedback with gradually changing magnitudes on the body, opening new possibilities for enhancing VR immersion.

\begin{acks}
We extend our gratitude to Singh Sandbox and its manager, Gordon Crago, for their generous support and provision of a wide array of tools, including electric hardware and lighting equipment. 
Additionally, we thank the reviewers for their insightful feedback, which has significantly improved our paper. ChatGPT is used in this work solely to correct spelling and grammar errors.
\end{acks}

\bibliographystyle{ACM-Reference-Format}
\bibliography{sample-base}

\end{document}